\documentclass{mn2e}
\usepackage[dvipsnames,usenames]{color}
\usepackage{colortbl}
\usepackage{times}
\usepackage{mathptmx}
\usepackage{graphicx}
\usepackage{amsmath}
\let\bibhang\relax

\usepackage{natbib}

\begin{document}
\title[3D MHD modelling of radio galaxy lobes]{Numerical modelling
  of the lobes of radio galaxies in cluster environments II: Magnetic
  field configuration and observability}
\author[M.J.\ Hardcastle \& M.G.H.\ Krause]
{M.J.\ Hardcastle$^1$ and M.G.H.\ Krause$^{2,3}$\\
$^1$ School of Physics, Astronomy and Mathematics, University of
  Hertfordshire, College Lane, Hatfield AL10 9AB\\
$^2$ Max-Planck-Institut f\"ur extraterrestrische Physik, Postfach
  1312, Giessenbachstrasse, 85741 Garching, Germany\\
$^3$ Excellence Cluster Universe, Technische Universit\"at M\"unchen,
  Boltzmannstrasse 2, 85748 Garching, Germany\\
}
\maketitle
\begin{abstract}
We describe three-dimensional magnetohydrodynamical modelling of
powerful radio galaxies in realistic poor cluster environments. This
modelling extends our earlier work on the hydrodynamics of radio
galaxies as a function of their cluster environment to consider the
magnetic field configuration in the lobes and its observational
consequences, using a realistic model for the magnetic field in the
intracluster medium, very high density contrast in the lobes and high
numerical resolution. We confirm, now with a realistic magnetic field
model, that lobes have characteristic trajectories in the radio
power/linear size diagram which depend strongly on their environment.
We investigate the detailed evolution of polarized emission, showing
that the lobes evolve from the initially ordered field configuration
imposed by our boundary conditions to one in which the longitudinal
field comes to dominate. We obtain simulated observations of
polarization whose properties are quantitatively consistent with
observations. The highly spatially intermittent magnetic field also
reproduces the observation that inverse-Compton emission from lobes is
much smoother than synchrotron. Our simulations allow us to study the
depolarizing effect of the external medium on the lobes, and so to
demonstrate that Faraday depolarization from environments of the type
we consider can reproduce the integrated fractional polarization
properties of large samples and the observed preferential
depolarization of the receding lobe.
\end{abstract}
\begin{keywords}
hydrodynamics -- galaxies: active -- galaxies: jets -- galaxies:
magnetic fields
\end{keywords}

\section{Introduction}
\label{intro}

Numerical modelling of powerful radio-loud active galaxies is an
essential tool towards developing a full understanding of the
evolution of these sources and their interaction with their
environment. On the large scales on which these sources are best
observed (physical scales of hundreds of kpc to Mpc) observations of
the best-understood class, powerful \cite{Fanaroff+Riley74} class II
(FRII) sources, show very complex, disparate structures in the radio
\citep[e.g.][]{Black+92,Bridle+94,Leahy+97} and also occupy a wide
range of hot gas environments, often also showing complex structure
\citep{Hardcastle+01,Wilson+06,Kraft+06,Kraft+07,Belsole+07,Chon+12,Ineson+13}.
Numerical modelling can now routinely deal with reasonable
approximations to the sort of large-scale hot-gas environments that
radio sources inhabit
\citep{Reynolds+02,Basson+Alexander03,Zanni+03,Krause05,Heinz+06,Mendygral+12}
as well as to smaller-scale galactic environments
\citep{Gaibler+11,Wagner+Bicknell11,Gaibler+12,Wagner+12,Cielo+14}
while the physics of the radio lobes can be made closer to what we
know to be realistic by considering magnetohydrodynamic (MHD) rather
than purely hydrodynamic (HD) modelling, and by considering electron
acceleration and loss processes
\citep{Clarke+89,Matthews+Scheuer90,Tregillis+01,Tregillis+04,Gaibler+09,ONeillJones10,Mignone+10,Huarte-Espinosa+11a}.
While numerical models never include all of the physics that is
necessary to describe a real radio galaxy, and are often limited by
numerical resolution, we may be converging on approximations that are
good enough to allow us to learn about the radio galaxy {\it
  population}, rather than individual pieces of radio galaxy physics,
by comparison with simulations.

In a previous paper \citep[][hereafter Paper I]{Hardcastle+Krause13}
we described two-dimensional, purely hydrodynamical models of the
radio-emitting lobes driven into groups and clusters of galaxies by
powerful double radio galaxies. Our objective in that paper was to
establish the general effect of such sources on the intracluster or
intragroup medium by studying the effect of varying (realistic)
parameters describing the medium on the radio source, and at the same
time to establish how well the time evolution of basic parameters of
the radio source (length, volume, synchrotron luminosity) matched
observations on the one hand and analytical models such as those of
\cite{Kaiser+Alexander97} on the other. Restricting ourselves
primarily to two-dimensional, axisymmetric modelling gave us
sufficient spatial resolution to model the jets from the kpc-scale on
which initially conical jets become collimated out to the scale of a
few hundred kpc on which large double radio sources are seen. We
established a number of important results, including (i) clear
evidence for departures from self-similarity of the lobes and for the
removal of the lobes from the centre of the cluster by buoyancy once
the inner parts of the lobes come into pressure balance with the
external environment, (ii)
evidence that tracks in the power-linear size diagram for powerful
radio sources are strongly influenced by their environment, and (iii)
evidence that a roughly constant fraction, close to unity, of the
energy stored in the lobes goes into work done on the external medium.

In the present paper we apply the same approach to three-dimensional,
magnetohydrodynamical (MHD) simulations of radio galaxies. Our reasons
for this are threefold:
\begin{enumerate}
\item It is important to check whether the conclusions of Paper I are
  valid in the case of a non-axisymmetric, magnetized outflow; earlier
  work (e.g. that of \cite{ONeill+05} and \citealt{Gaibler+09} for MHD) supports this, but the best way
  of testing it is to run more realistic simulations in the same
  environments as those of Paper I.
\item Magnetic fields, in particular, are essential for realistic
  synchrotron visualization, and allow us to consider polarized
  emission from our simulated radio sources for the first time. This
  allows comparison with the work of
  \cite{Huarte-Espinosa+11a,Huarte-Espinosa+11}, repeating their
  pioneering work in this area at higher numerical resolution and
  comparing with observations.
\item Incorporating three-dimensional structures and magnetic fields
  is a further step on the way towards models of FRII radio galaxies
  that contain all the physics necessary to compare with observations.
  Here, we aim to establish the magnetic field configuration in radio
  lobes.
\end{enumerate}

Section \ref{sec:setup} describes the setup of our new simulations and
Section \ref{sec:setup} our results. The key results and the extent to
which we believe them to be relevant to real objects are summarized in
Section \ref{sec:discussion}.

\section{The simulation setup}
\label{sec:setup}

Our modelling in this paper, as in Paper I, uses the MHD
implementation in the freely available code
PLUTO\footnote{http://plutocode.ph.unito.it/}, version 4.0,
described by \cite{Mignone+07}. In particular, we used the {\tt hlld} solver
with the divergence cleaning algorithm to enforce $\nabla \cdot
{\mathbf B}=0$ (the eight-wave method tended to wash out fine magnetic
structure in the simulated lobes while we were unable to use
constrained transport with our choice of internal boundary condition).
We used dimensionally unsplit second-order Runge-Kutta time stepping
with a Courant-Friedrichs-Lewy (CFL) number of 0.3.

Since in this paper we are attempting to carry out fully
three-dimensional, MHD calculations, we cannot use the very high
resolution of Paper I, which was intended to allow us to generate
self-consistent, collimated jets from an initially biconical outflow.
Time and physical memory constraints restrict us to a fiducial
simulation using a $400\times 400\times 400$-element volume. We do not
use the adaptive mesh refinement capability of PLUTO in order to
preserve the small-scale magnetic field fluctuations in the jet's
ambient medium (see below for details). As a direct consequence of
this grid setup, we cannot inject a conical jet which will
self-collimate on the appropriate physical scale -- instead we take a
similar setup to that used by \cite{Huarte-Espinosa+11a} (hereafter
HE11), and inject a bipolar jet using an internal boundary condition which is
a cylinder of radius $r_j$ and length $l_j$ along the $x$-axis,
internal to which we have $\rho = \rho_j$, $v_x = \pm{\cal M}_jc_s$, $v_y
= v_z = 0$ and $p = T_j \rho_j$. The injected field structure is
purely toroidal with $|{\mathbf B}| = B_j$. A conserved tracer
quantity is injected with the jets, taking the value 1.0 at injection
and 0 everywhere else in the grid at the start of the simulation. Both
the injected jets and the ambient medium are treated as ideal gases
with $\gamma = \frac{5}{3}$. The outer boundary
conditions are periodic.

\begin{table*}
\caption{Key initial simulation parameters in physical units}
\label{physunits}
\begin{tabular}{llrll}
\hline
Fluid&Quantity&Value&Units&Location\\
\hline
Jet&Kinetic power&$10^{38}$&W&(One-sided)\\
&Density&3.75&m$^{-3}$&Injected\\
&Speed&$7.3 \times 10^4$&km s$^{-1}$&Injected\\
&Temperature&8&MeV&Injected\\
&Radius&4.2&kpc&Injected\\
&External Mach number&100&&Injected\\
&Internal Mach number&1.6&&Injected\\
&Magnetic field strength&0.15&nT&Injected, fiducial\\
&Plasma $\beta$&400&&Injected\\
Cluster&Density&$3 \times 10^4$&m$^{-3}$&Central\\
&Temperature&2.0&keV&Everywhere\\
&Sound speed&730&km s$^{-1}$&Everywhere\\
&RMS field strength&0.7&nT&Central\\
&Plasma $\beta$&40&&Everywhere\\
\hline
\end{tabular}
\end{table*}

For continuity with Paper I, we take the environments of the radio
sources to be rich groups or clusters, with a density profile given by
\begin{equation}
n = n_0\left[1+\left(\frac{r}{r_{\rm c}}\right)^2\right]^{-3\beta/2}
\label{eq:betamodel}
\end{equation}
The core radius and value of $\beta$ are variable, as described
  below. PLUTO simulations are run using scalable simulation units,
  but in our case we wish to represent a specific physical situation,
  so, to match Paper I, we take the simulation unit length to be $L =
2.1$ kpc, the unit temperature to be $kT = 2$ keV, and the central
density to be $n_0 = 3 \times 10^4$ m$^{-3}$. As in Paper I $c_s =
730$ km s$^{-1}$, meaning that the simulation time unit is $\tau =
L/c_s = 2.9 \times 10^6$ years. In terms of volume, we go out to a
maximum radius of $\pm 150$ simulation units ($\pm 310$ kpc), again as
in Paper I, so that the volume elements of our fiducial simulation are
0.75 simulation units, or 1.6 kpc on a side.

The assumptions we make about the environment then to some extent
control the properties of the jet that is injected. We must put in by
hand the condition that was automatically satisfied by our
self-collimating jets in Paper I, namely that $p_j \approx p_0$, where
$p_0$ is the central pressure, or $p_j \approx 1$ in simulation units.
Thus if the jet is light, with $\rho_j \ll 1$ in simulation units,
then necessarily $T_j \approx 1/\rho_j$; we follow Paper I, \cite{Gaibler+09}
and ultimately \cite{Krause03} in our assumption that a very light jet
is required to make realistic lobes. If we require that the initial jet
advance be supersonic, then in addition ${\cal M}^2_j\rho_j > 1$. Note
that marginally satisfying this constraint, with ${\cal M}^2_j\rho_j =
1$, leads to $T \approx {\cal M}_j^2$, i.e. to a jet which is
trans-sonic in terms of its internal sound speed at injection.
However, since the nominally injected jet power can be written (in
simulation units, and writing in
terms of the enthalpy since the jet is hot, we have
\begin{equation}
Q = \pi r_j^2 {\cal M}_j \left(\frac{1}{2} {\cal M}_j^2\rho_j + \frac{5}{2}
T_j\rho_j + \frac{B^2}{2}\right)
\end{equation}
we see that the jet that supplies minimal power for a given $r_j$
while still obeying the constraint ${\cal M}^2_j\rho_j > 1$ is
precisely this type of trans-sonic, hot jet, and in this limit the
energy supply is predominantly in the form of the enthalpy of the jet
material rather than the kinetic energy density (assuming that the
term involving the magnetic energy density may be neglected). We can
change the balance between kinetic and thermal energy content only by
increasing the jet Mach number such that ${\cal M}_j^2 \gg 1/\rho_j$. 

For our simulation setup, we have a further constraint which is that
$r_j$ cannot be made so small that it is not reasonably well resolved
by the simulations (i.e. smaller than a few simulation cells). We
adopt ${\cal M} = 100$, and then choose $\rho_j = 1.25 \times
10^{-4}$, $T=4000$ and $r_j = 2.0$ in simulation units: this gives a
(one-sided) jet power in physical units of $10^{38}$ W, consistent
with our previous work (see Table \ref{physunits} for the full
physical properties of the jet). The external Mach number of 100 (corresponding
to a speed around $0.24c$ in physical units) is the largest that can
reasonably be used without needing to take account of the effects of
special relativity, and the jet temperature corresponds to
relativistic particle energies for electrons, so that we are at least
close to the regime thought to be occupied by real radio galaxies.
However, it is important to note that the jet is slower than, and
  very likely has a lower internal Mach number than, the jets in real
  FRII radio galaxies; a consequence of this is likely to be weaker
  internal shocks and a consequent suppression of `hotspot'
  structure. We comment further on the relationship between our
  assumed source properties and real FRII sources in Section
\ref{sec:critik}.

\begin{figure*}
\includegraphics[width=0.88\linewidth]{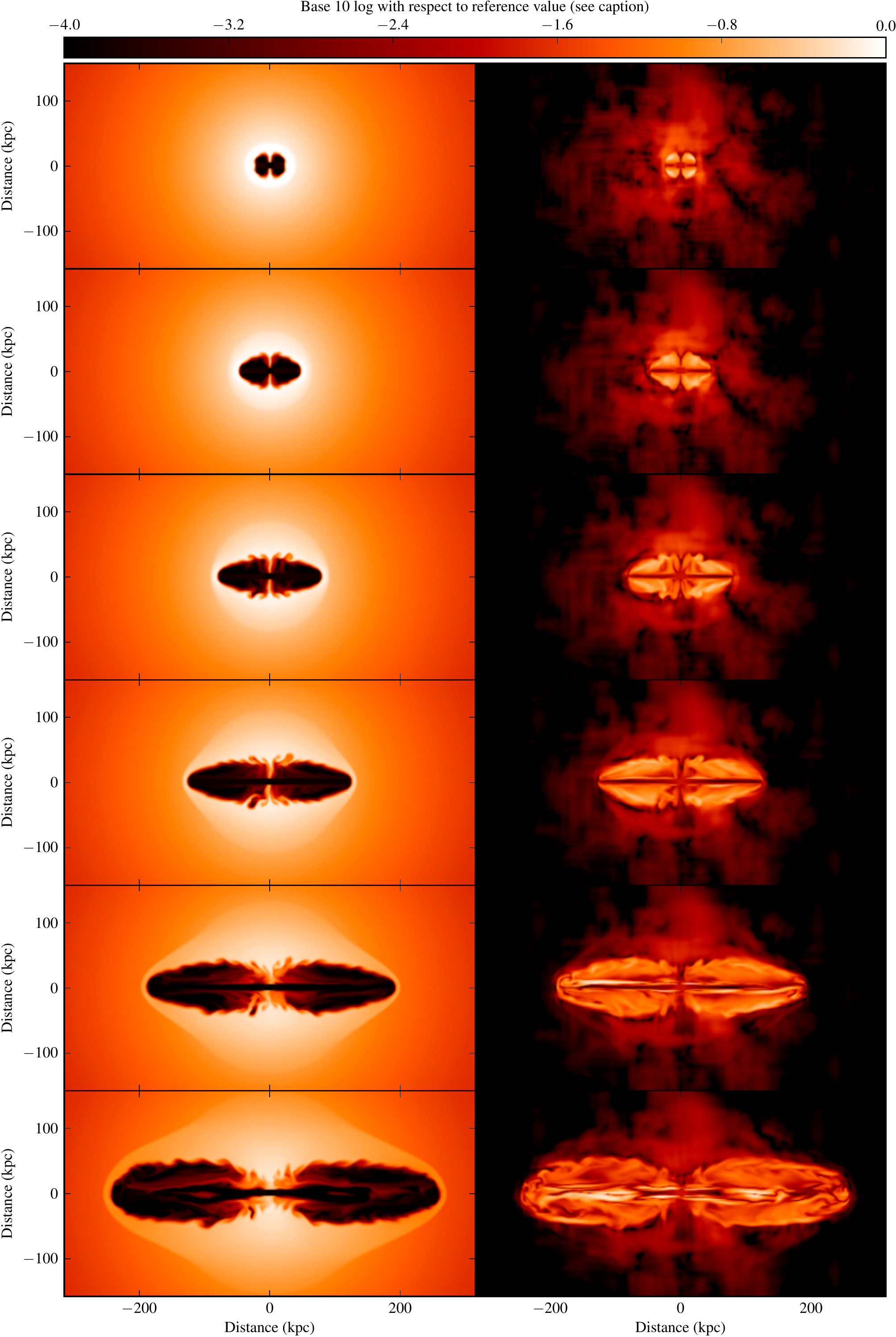}
\caption{Particle density (left) and magnetic field energy density (right) slices in the
  central $xy$ plane for B75-30 at simulation times $t=30, 60\dots
  180$ Myr as a function of position. Colours show the base-10
  logarithm of both quantities relative to the cluster-central
    value at $t=0$ (for density) and a reference value corresponding
    to $B=1.8$ nT (for field energy density).}
\label{fig:slices}
\end{figure*}

We adopt a fiducial value of $B_j$ of 0.05 (simulation units) which
corresponds to about 0.15 nT in physical units. This input number constrains
the normalization of the field in the lobes, as we will discuss later.
More importantly, we follow HE and set the field in the ambient medium
to be a Gaussian random field whose energy density scales with
pressure, with a peak field strength in the centre of the medium of
0.7 nT (this is consistent with field strength measurements in real
groups and clusters of galaxies, see e.g. \citealt{Guidetti+12}). We
put in a power spectrum of the magnetic field strength corresponding
to Kolmogorov turbulence in the manner described by \cite{Murgia+04}
and \cite{Hardcastle13}, cutting the spectrum off below size scales of
$\sim 3$ pixels to avoid injecting high-spatial-frequency structure
into the simulations (this would immediately be damped out by
numerical effects in any case, and removing it gives a more useful
baseline simulation at $t=0$). We emphasise that the scaling of
  the magnetic energy density with pressure ensures that little power
  is present in the spectrum of the magnetic field on scales larger
  than the core radius, though the unscaled power-law spectrum extends
  to structures with scales comparable to the half-size of the
  simulation volume. In this way we attempt to model a
realistic magnetic environment for the jet.

We run three sets of simulations to investigate the jet behaviour as a
function of different parameters:
\begin{enumerate}
\item A set of 9 simulations varying the parameters of the
  environment, with $\beta = 0.55$, 0.75 and 0.90 and $r_c = 20$, 30
  and 40 simulation units (roughly 40, 60 and 80 kpc). This is similar
  to the range studied in Paper I except that in this paper we do not carry out
  simulations with $\beta = 0.35$, which we consider rather too flat
  to be realistic.
\item A set of 4 simulations varying the input magnetic field
  strength, all with $\beta = 0.75$ and $r_c = 30$. The magnetic field
  strength injected in the jet controls the field strength found in
  the lobes. In simulation units these have $B_j = 0.01$, 0.02, 0.05 and
  0.1.
\item A set of two simulations with $\beta = 0.75$, $r_c = 30$, $B_j =
  0.05$, one of which has a significantly higher resolution (a $600
  \times 300 \times 300$ simulation covering the same physical volume
  in the $x$ direction, i.e. with elements a factor 1.5 smaller in all
  dimensions). This allows us to investigate any resolution effects in
  our results on e.g. polarization. Because of the smaller physical
  size of the simulations in the $y$ and $z$ directions, the
  high-resolution simulation does not capture the full shock size at
  late times, and so needs to be used with caution when the external
  medium is considered. These two simulations have the same initial
  conditions for the external medium so that effects of different
  environments can be distinguished from resolution effects.
\item One simulation with $\beta = 0.75$, $r_c = 30$ with standard
  resolution but a $664 \times 300 \times 300$ box, thus allowing the
  lobes to extend to $\pm 250$ simulation units at the cost of no
  longer being able to track the shocked gas accurately. This run is
  only used where required to test convergence of lobe properties that
  evolve with length.
\end{enumerate}

In what follows we assign each simulation a code of the form Bxx-yy...
for brevity; these codes are used to identify results from individual
simulations in the plots in the following section. The codes and the
properties of the simulations they map onto are listed in Table
\ref{tab:summary}. As the B75-30 simulation (and its high-resolution
counterpart B75-30-HR) have non-extreme environmental properties they
are generally used as illustrations when specific simulations are
required in what follows: their properties can be taken as
representative unless otherwise stated.

\begin{table}
\caption{Parameters of the ambient density profile, $\beta$ and
    $r_c$ as in eq.\ \ref{eq:betamodel} explored in this paper, with other specifics of the
individual simulations. Cluster core radii are given in simulation units,
equal to 2.1 kpc.}
\label{tab:summary}
\begin{tabular}{lrrl}
\hline
Code&$\beta$&$r_c$&Other notes\\
&&(sim.\ units)&\\
\hline
B55-20&0.55&20&\\
B55-30&0.55&30&\\
B55-40&0.55&40&\\[2pt]
B75-20&0.75&20&\\
B75-30&0.75&30&\\
B75-40&0.75&40&\\[2pt]
B90-20&0.90&20&\\
B90-30&0.90&30&\\
B90-40&0.90&40&\\[2pt]
B75-30-HM&0.75&30&$2\times$ normal jet field\\
B75-30-LM&0.75&30&$0.4\times$ normal jet field\\
B75-30-VLM&0.75&30&$0.2\times$ normal jet field\\[2pt]
B75-30-HR&0.75&30&$1.5 \times$ normal resolution\\[2pt]
B75-30-LONG&0.75&30&Long, thin grid\\
\hline
\end{tabular}
\end{table}

All our production runs were carried out using the STRI cluster of the
University of Hertfordshire, using between 128 and 256 Xeon-based
cores, typically for about 24-48 hours per run, terminating the run
when the shocked region reaches the grid boundary. (Simulations other
than B75-30-LONG therefore all extend to a total lobe length of around
300 kpc, though the amount of simulation time required to
reach the edge varies widely.) PLUTO was configured to write out the
complete state of the simulation every 1 simulation time unit (3 Myr) and
these images of the simulation grid were used to compute derived
quantities such as the dimensions of the lobe, the energy stored in
the lobe and in the shocked region, and so forth. We use the conserved
tracer quantity to define regions inside and outside of the lobes in
the analysis that follows. For consistency with Paper I, we define a
contiguous volume as being inside the lobes, which in this paper is
the volume internal to a surface at which the tracer quantity is
everywhere $>10^{-3}$. (We note that it is possible for the tracer to
be $<10^{-3}$ inside this volume, e.g. if there is large-scale
entrainment of thermal material, but this definition in practice
appears to give a sensible lobe boundary, as did the equivalent
definition in Paper I.) The results are insensitive to the tracer
threshold chosen as long as it is not too high. We define the shocked
region as being the region external to the lobe boundary with an
outward radial velocity greater than 0.1 in simulation units. The
volume outside the shocked region is not considered in any calculation
of the energetics. As in Paper I, calculation of energies for the
shocked region take as their zero point at any given time $t$ the
energy stored within the boundary of the shock at time $t$ in the
image from $t=0$, so that we correct for the pre-existing internal
energy of the environment.

In postprocessing we retain the ability to distinguish between the two
lobes of the radio galaxy (the magnitude of differences between the
two gives an estimate of the effects of the slightly different initial
conditions on the two sides of the source) but in most plots presented
below the quantities for the two lobes are summed or averaged, as
appropriate, so as to give a single curve for each run. In particular,
the reader should note that the lobe length, used in many of the plots
below, is actually the mean of the lengths of the two lobes.

\section{Results}

In this section and throughout the remainder of the paper all results
are presented converted to physical units, except where otherwise
noted, in order to facilitate comparison with observations.

\label{sec:results}

\subsection{General simulation behaviour}
\label{sec:general}
Fig.\ \ref{fig:slices} shows density and magnetic field energy density
slices through the central $xy$ plane of the B75-30 simulation run,
illustrating the broad features of the simulations' behaviour.
Qualitatively we see very similar behaviour to what was observed in
Paper I in many respects, as expected since the objects we are
simulating and their environments should be very similar. Here we
focus on the important differences between the two sets of
simulations.

\begin{itemize}
\item The expansion in the new simulations is slower at early times
  -- this is because lobe formation is immediate in the new
  simulations as a result of the pre-heating of the jet material. The
  small aspect ratio of the lobes at early times in the new
  simulations (e.g.\ top panel of Fig.\ \ref{fig:slices}) is probably not realistic, however.
\item The gap between the lobes is present from the earliest times in
  the new simulations, as a consequence of the non-zero starting
  length of the jet -- it does not emerge naturally from the
  simulations as in Paper I. However, its growth with time is
  qualitatively consistent with our earlier work.
\item The shape of the shock driven into the external medium is
  different at all times -- presumably as a consequence of the
  different lobe dynamics.
\end{itemize}

\begin{figure*}
\includegraphics[width=0.9\linewidth]{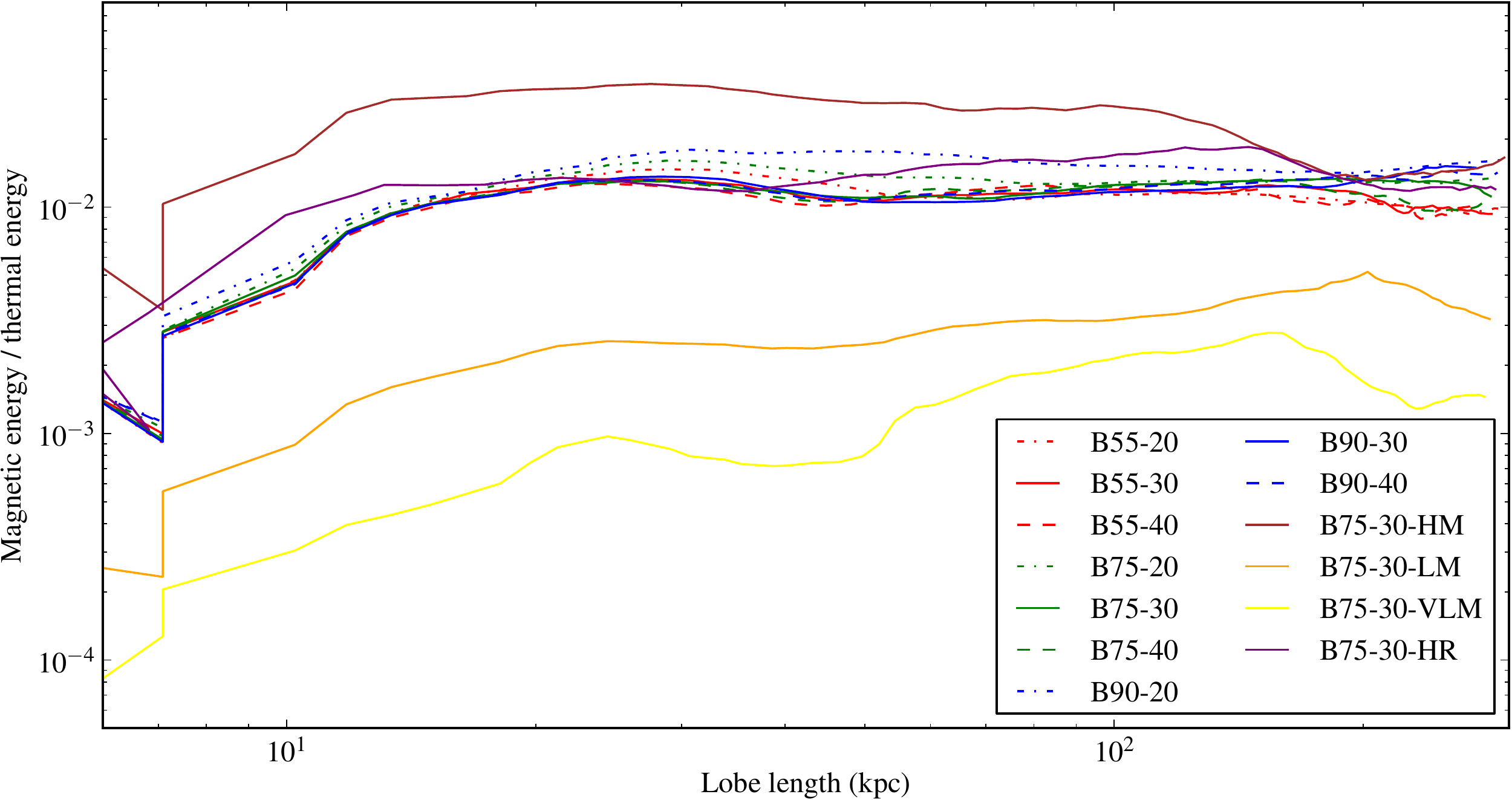}
\caption{The ratio between the energy in field and that in particles
  integrated over the lobes, as a function of source length.}
\label{fig:bfield}
\end{figure*}

Another important difference is of course the presence of magnetic
fields, as shown in the right-hand panel of Fig.\ \ref{fig:slices}.
The input magnetic field strength is chosen so as to mean that the
magnetic field is globally dynamically unimportant, though locally
that may not be the case; for example, the field is higher at the lobe
edges and may provide non-negligible tension there. The global picture is illustrated in
Fig.\ \ref{fig:bfield}, which shows the total energy stored in
magnetic field in the lobes compared to that stored in particle
pressure as a function of lobe length. We see that the initial input
field strength affects this, but that it is otherwise essentially
independent of lobe length, environment, and numerical resolution. A
mean ratio in energy density $U_B/U_e \sim 0.01$ (mean lobe plasma $\beta
\approx 100$) is consistent with inverse-Compton energy estimates in
real radio sources \citep{Kataoka+Stawarz05,Croston+05-2}.

\begin{figure*}
\includegraphics[width=\linewidth]{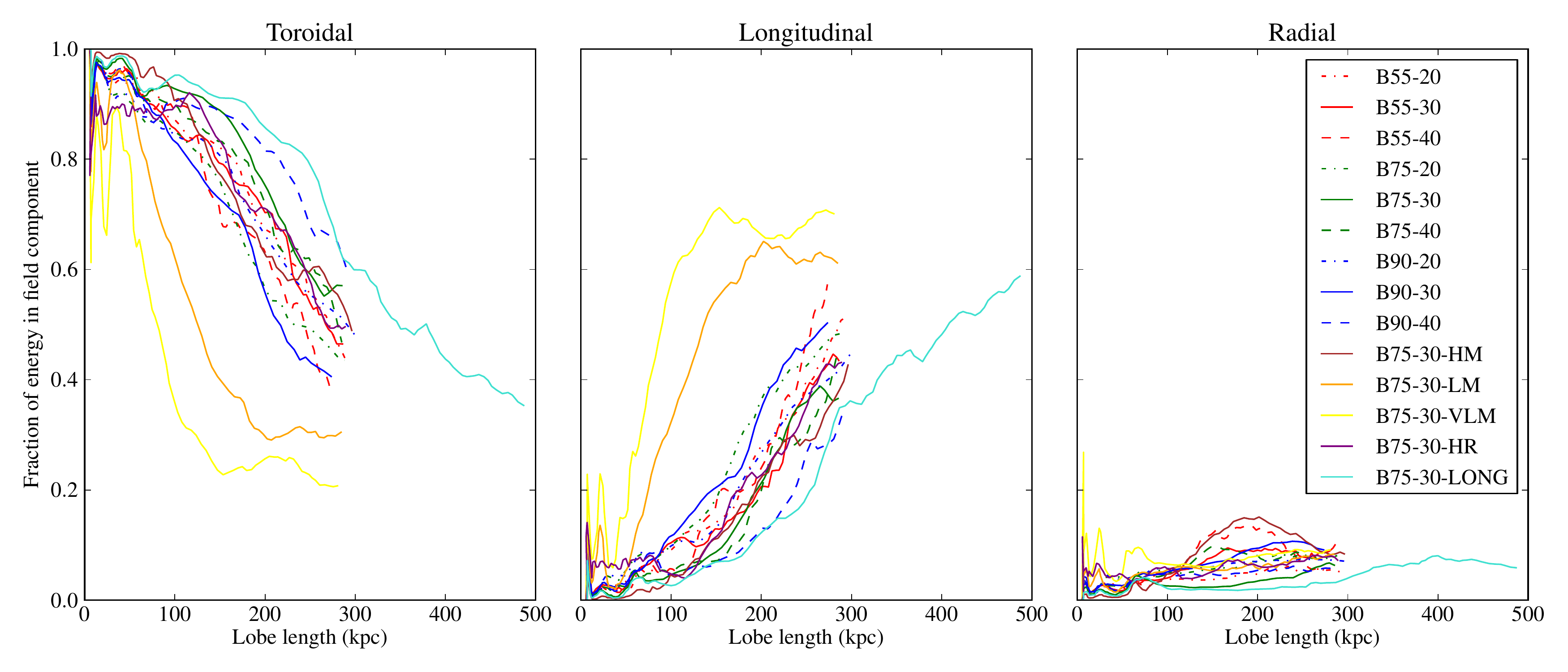}
\caption{The fraction of magnetic energy in (left) toroidal, (middle)
  longitudinal and (right) radial components of the magnetic field.
  The long simulation B75-30-LONG is plotted here to show the
  long-term evolution of the field.}
\label{fig:bcomp}
\end{figure*}

The simulations also give us the overall direction of the magnetic
field in all times, which we may describe in terms of its toroidal,
longitudinal and radial components relative to the jet axis (the
$x$-axis). We find that there is always initially a strong toroidal
component of the field, in the same sense as the injected toroidal jet
field. At later times (earlier with weaker fields or at higher
numerical resolutions), there are significant longitudinal and radial
components as well, reflecting the dynamical effects of large-scale
bulk motions inside the lobes (Fig.\ \ref{fig:bcomp}), and the
longitudinal component becomes steadily more important as the lobes
grow, possibly saturating around 60 per cent of the total magnetic
field energy density as we see in the case of the weak-field runs
B75-30-LM and B75-30-VLM. The evolution of this field component
affects the polarization behaviour of the lobe, as we shall see in
later sections.

\subsection{Lobe dynamics}
\label{sec:dynamics}

\begin{figure*}
\includegraphics[width=0.8\linewidth]{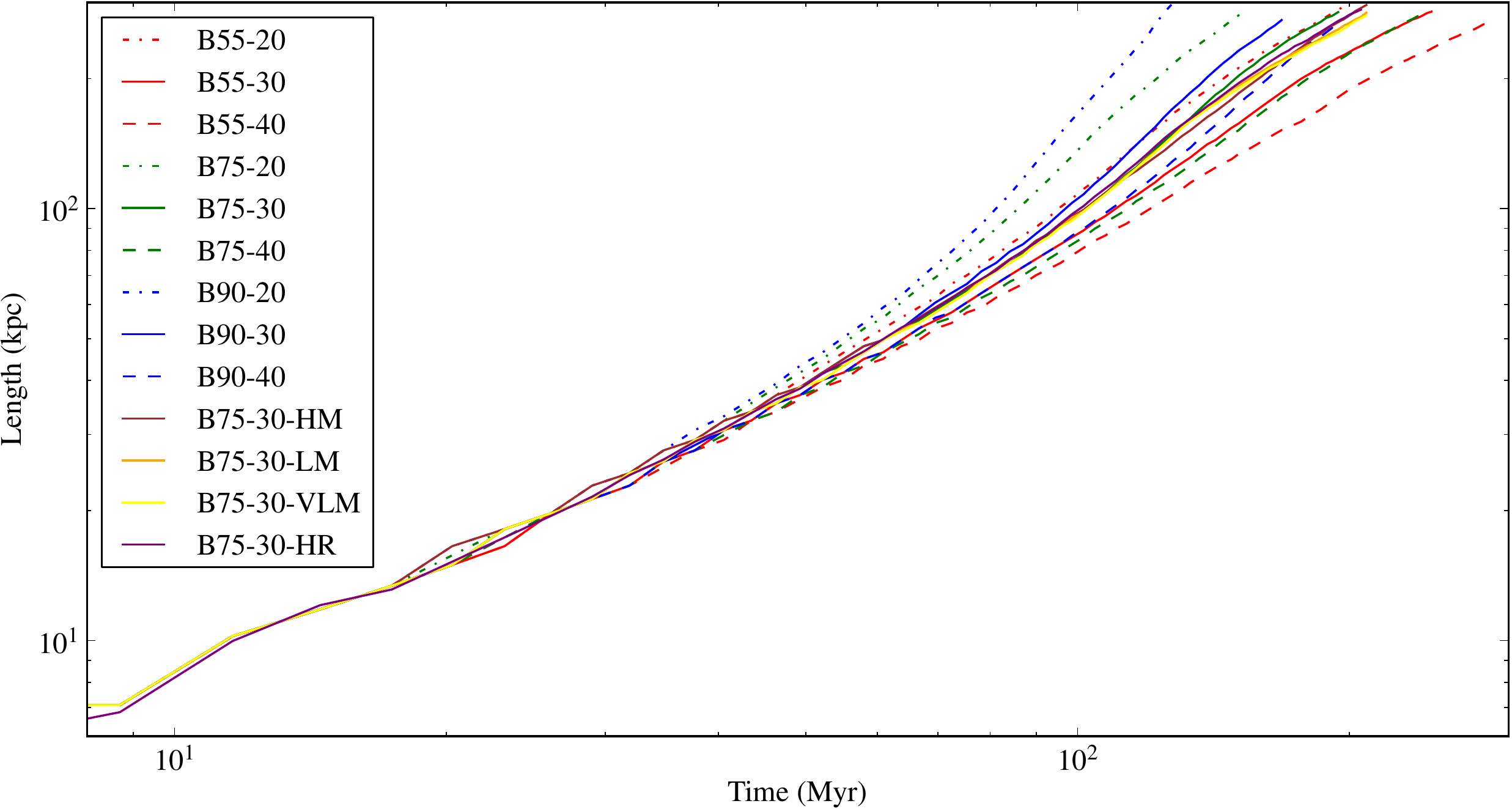}
\includegraphics[width=0.8\linewidth]{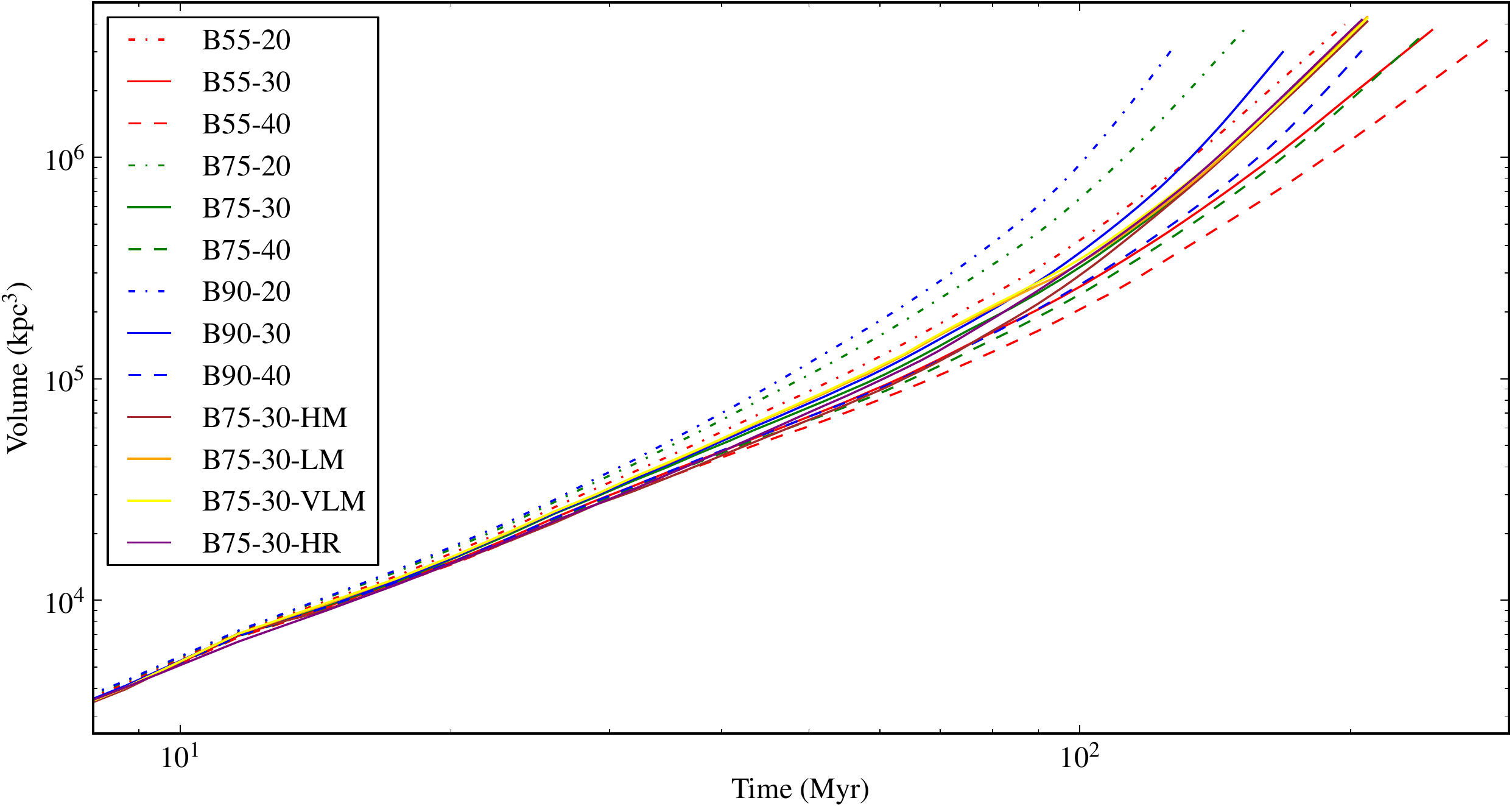}
\includegraphics[width=0.8\linewidth]{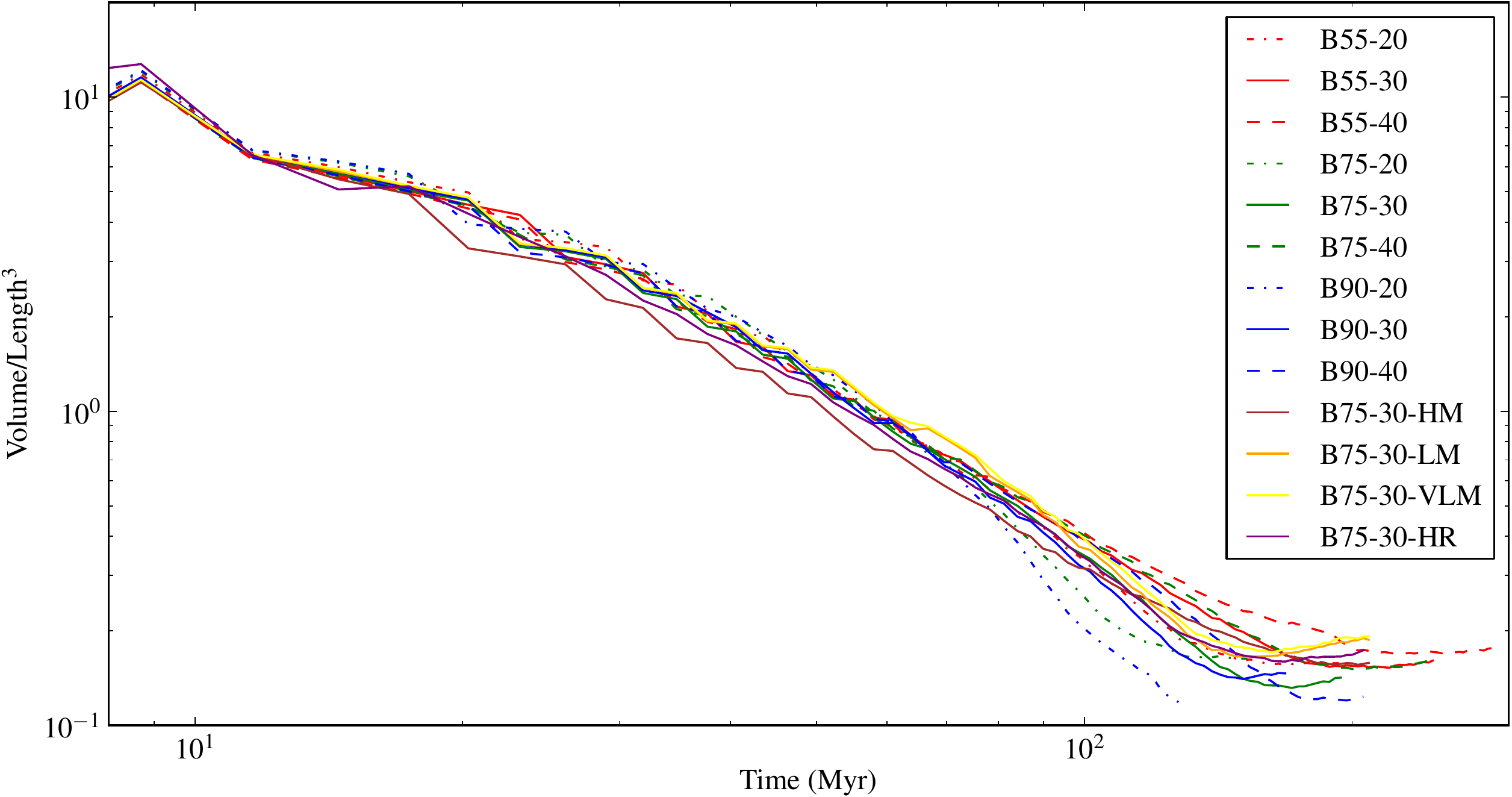}
\caption{Growth of the lobes with simulation time. Top panel: mean
  lobe length $L$.
  Middle panel: mean lobe volume $V$. Bottom panel: $V/L^3$.}
\label{fig:dynamics}
\end{figure*}

Quantitatively, the lobes show the expected approximately linear
growth with time at early times; the lobe growth rate is limited by
ram pressure balance at the jet head and so is roughly constant within
$r < r_c$. Beyond the core radius of the environment, we see the
expected steepening -- the predicted slope \citep{Kaiser+Alexander97}
is $5/(5-3\beta)$, and the three $\beta$ values clearly show the
expected trends. The lobe volume scales with lobe length, but not in
the sense expected from self-similarity: this is shown by the bottom
panel of Fig.\ \ref{fig:dynamics}, which shows that the axial ratio
varies by roughly an order of magnitude over the simulation time. The
value plotted here would be constant if the lobes were self-similar.
Departure from self-similarity is expected in real sources
\citep{Hardcastle+Worrall00}, as discussed at length in Paper I, and
is a general feature of simulations (e.g. \citealt{Krause05}) so long
as the source is not strongly overpressured at all times.

It is important to note at this point that the simulations with
varying magnetic field and resolution do not deviate significantly
from the fiducial simulation with the same environmental parameters.
This is as expected, and shows that neither resolution nor magnetic
field strength in the lobes affects the overall source dynamics.

\subsection{Energetics of the lobe and environmental interaction}
\label{sec:energetics}

\begin{figure*}
\hskip 10pt
\includegraphics[width=0.45\linewidth]{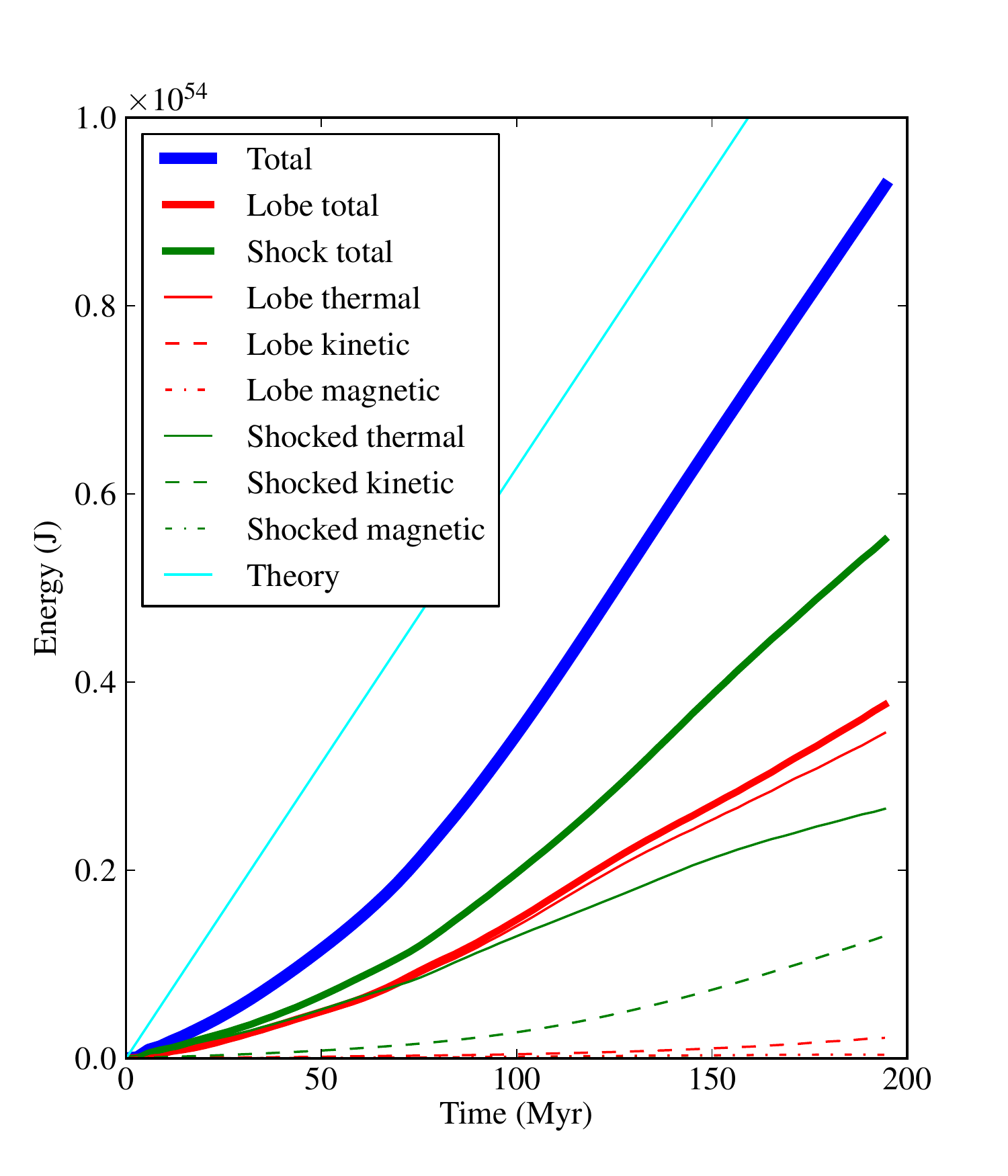}
\includegraphics[width=0.45\linewidth]{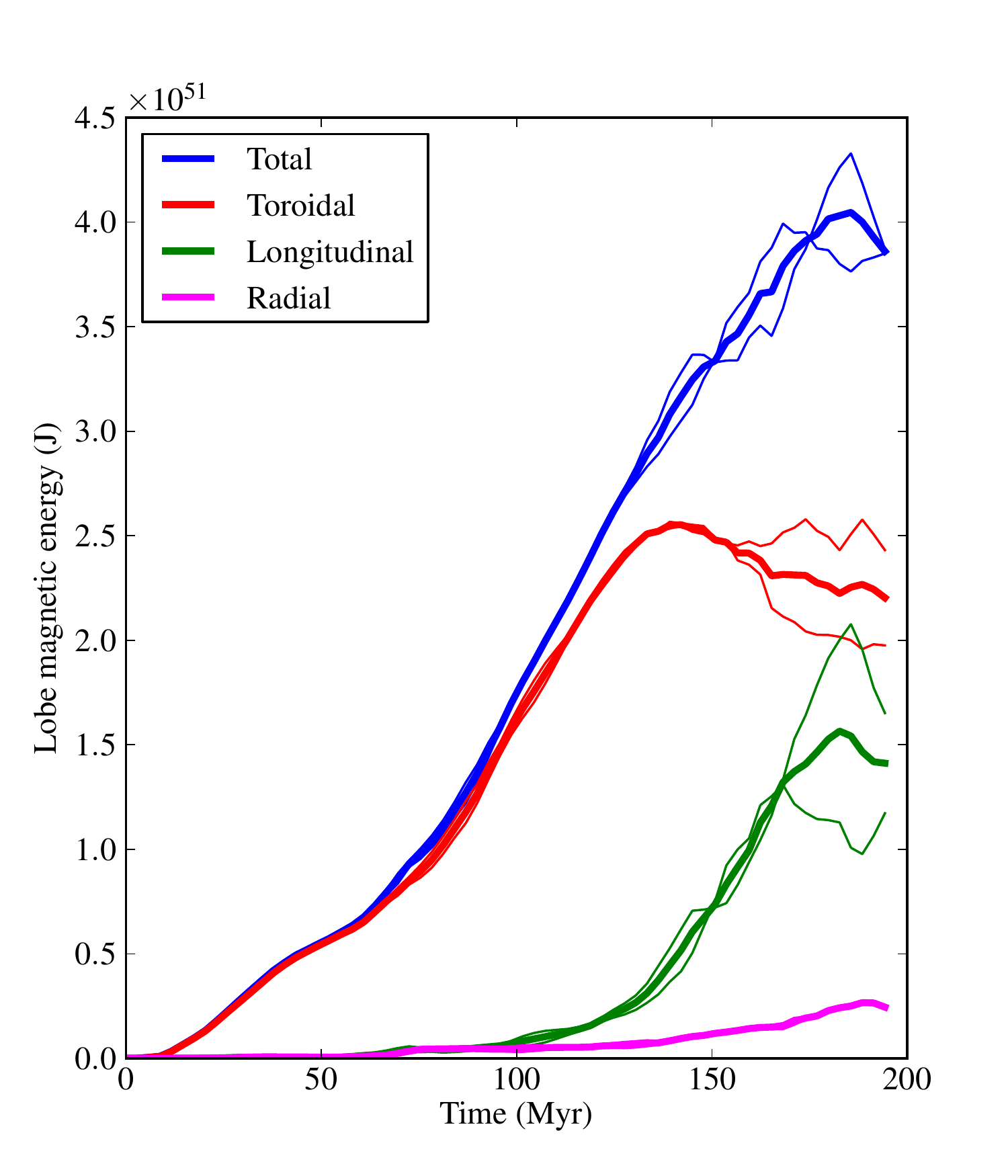}
\caption{Left: energies in the lobe and shocked region as a function
  of simulation time for the B75-30 simulation. Also plotted
  (`theory') is the expected growth of the total if a constant energy
  flows in through the jet boundary condition. Potential energies are
  calculated in post-processing but not plotted here as their values
  are so small. Right: the same simulation, with the growth of lobe
  magnetic field energy and its three components in cylindrical
  co-ordinates as a function of simulation time. Thin lines on this
  plot show the curves for the two lobes separately (multiplied by 2)
  to provide a sense of the scatter imposed by `weather' in the
  simulations.}
\label{fig:energyinput}
\end{figure*}

The total thermal (particle), kinetic and magnetic energies in the
lobes and shocked region can be plotted as a function of time to
investigate the rate of energy injection into the system. An example
is shown in Fig.\ \ref{fig:energyinput} (left panel). We see that the
energy input into the system falls below the expected value at all
times: as with the simulations of Paper I, this arises because at
early times the jet boundary condition does not couple well to the
ambient medium, so that the jet flow on to the grid is partially
suppressed. Once the lobes are well established the gradient of total
energy is very comparable to expectations. The effect of this is to
artificially lower the jet power at early times, at worst by of order
a factor 2; this should be borne in mind when considering later
results. However, at late times the discrepancy in total energy is
only of order 25 per cent and can safely be ignored.

While the growth in the total energy in the system is linear, the
growth in individual components need not be, as shown in
Fig.\ \ref{fig:energyinput} (right panel). The total energy in
the magnetic field grows roughly as a constant fraction of the total
energy in the system, but the longitudinal and toroidal components behave quite
differently, reflecting the change in the field structure discussed
above (Section \ref{sec:general}). At about half-way through the run
energy in toroidal field ceases to grow and even decreases slightly,
with all subseqent increase being in the energy stored in the
longitudinal and (to a much lesser extent) radial components.

\begin{figure*}
\hskip 10pt
\includegraphics[width=\linewidth]{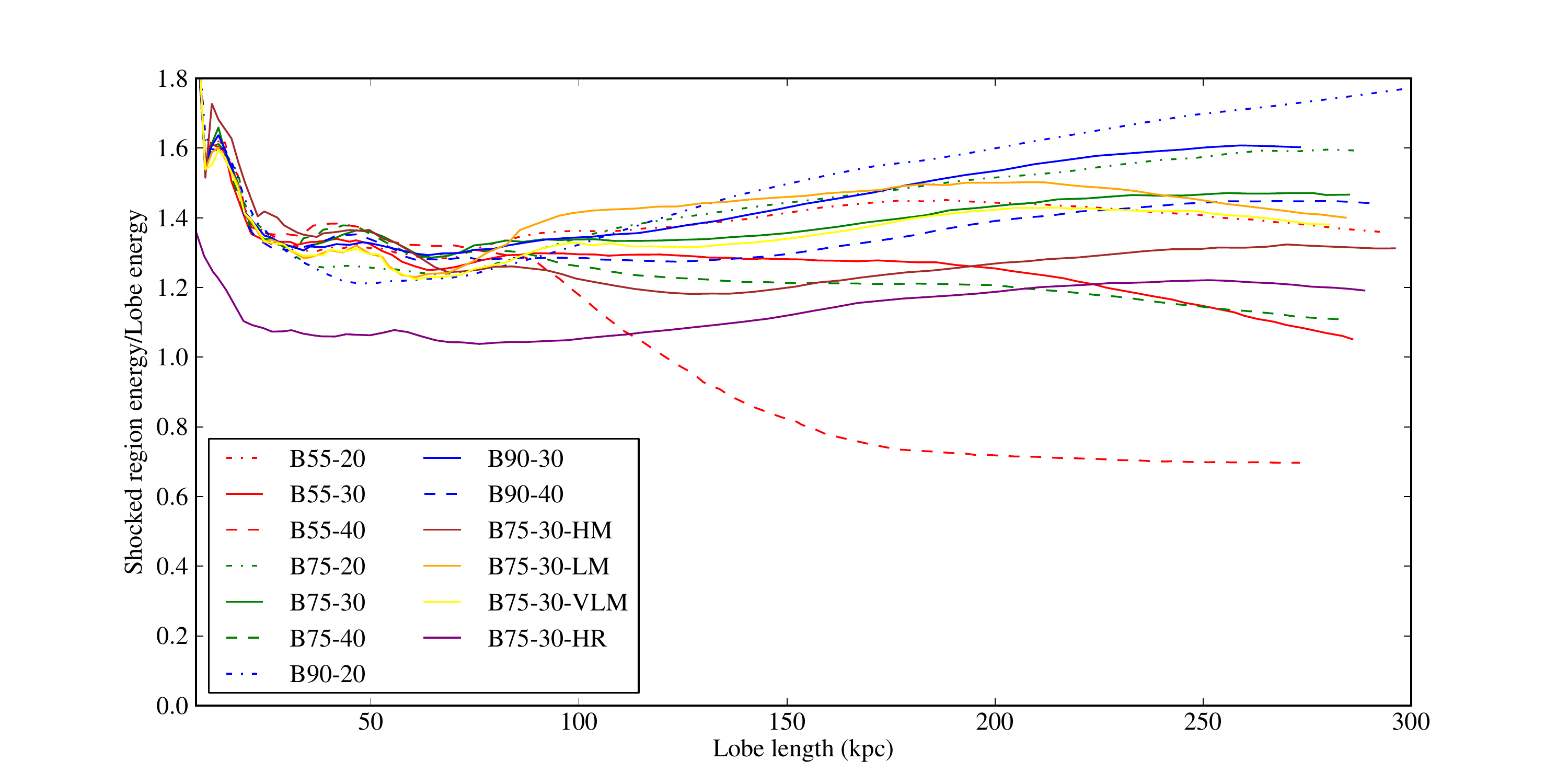}
\caption{The radio of the energy in the shocked external medium and in
the lobes as a function of lobe length.}
\label{fig:eratio}
\end{figure*}

We can also look at the ratio of energies stored in the shocked region
and the lobe. One of the key results of Paper I was that this ratio
was close to (typically a little larger than) unity, irrespective of
the environment in which the source expanded. Fig.\ \ref{fig:eratio}
shows this ratio as a function of lobe length. We see that in the
current simulations this ratio is in the range 0.6-1.8, consistent
with the previous work, but the scatter is a little larger than in
Paper I and there are clear environmental dependences in the sense
that the ratio is highest for higher $\beta$ and lower $r_c$, i.e. for
lower-mass environments. It is not clear whether this difference with
Paper I is because of the somewhat different lobe dynamics, as
discussed above, or whether it arises from the absence of the
naturally self-collimating jets of Paper I. However, the overall
general agreement with the conclusion of Paper I is encouraging.

\subsection{Integrated radio emission}
\label{sec:integratedradio}

\begin{figure*}
\includegraphics[width=0.8\linewidth]{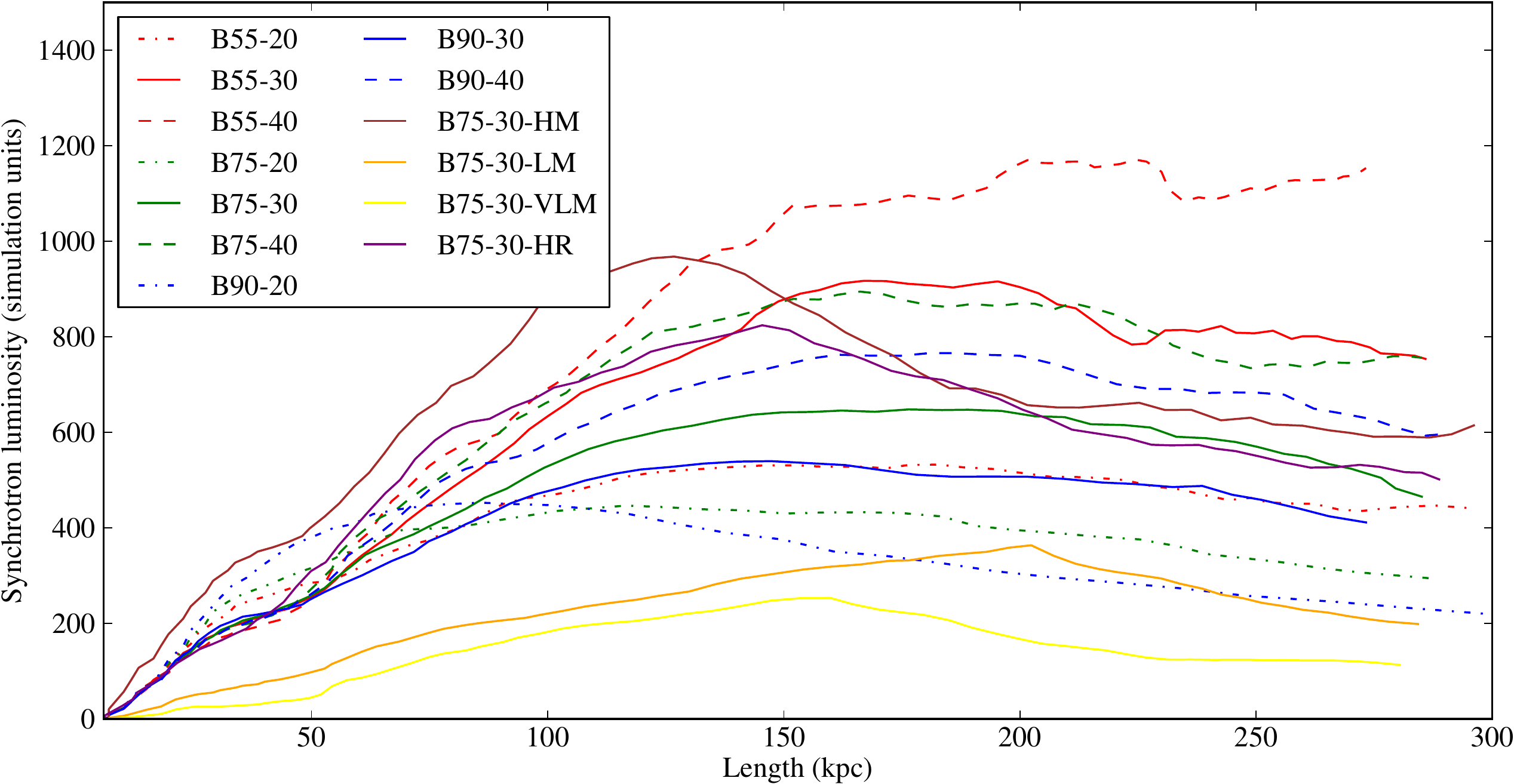}
\includegraphics[width=0.4\linewidth]{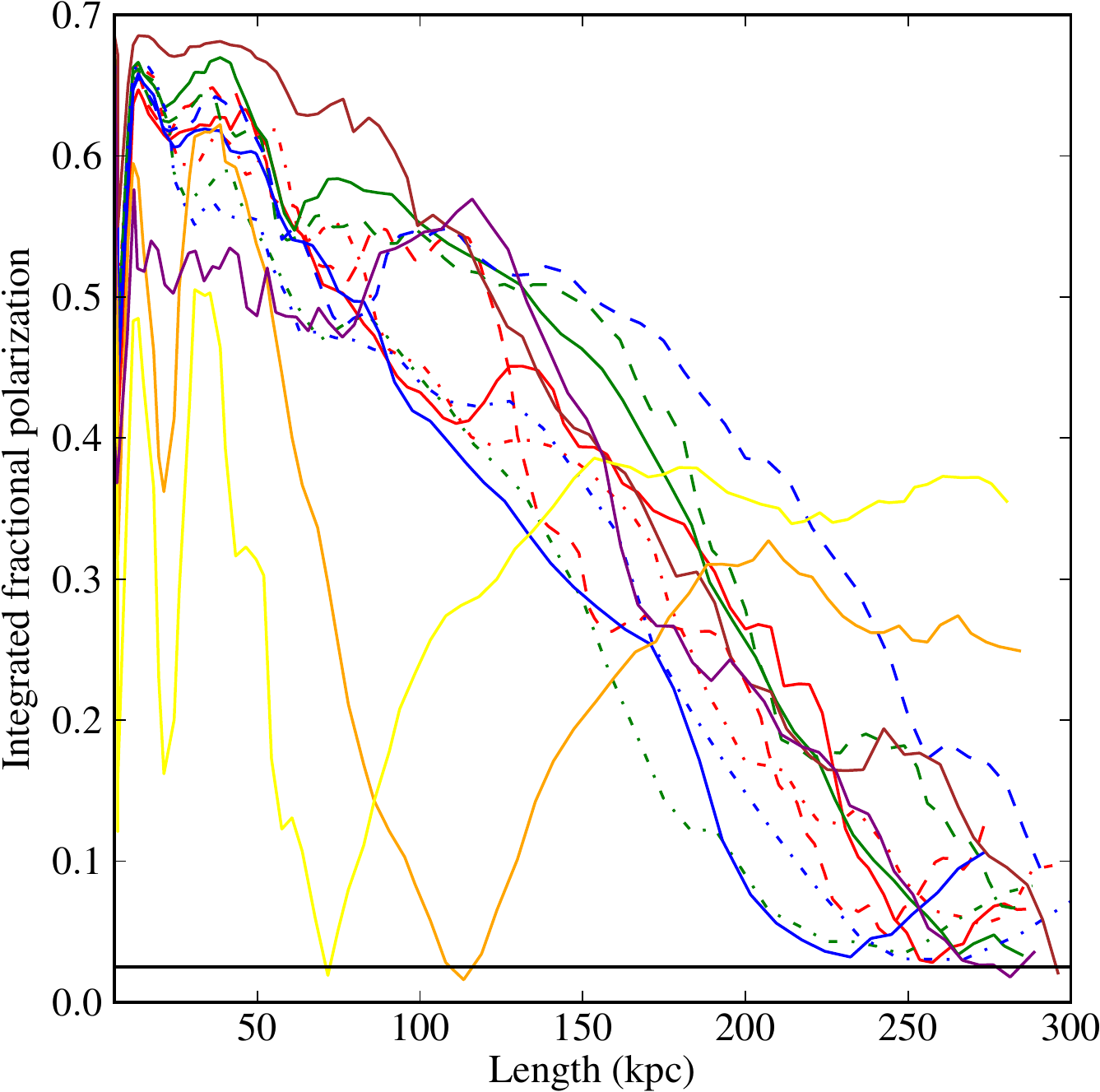}
\includegraphics[width=0.4\linewidth]{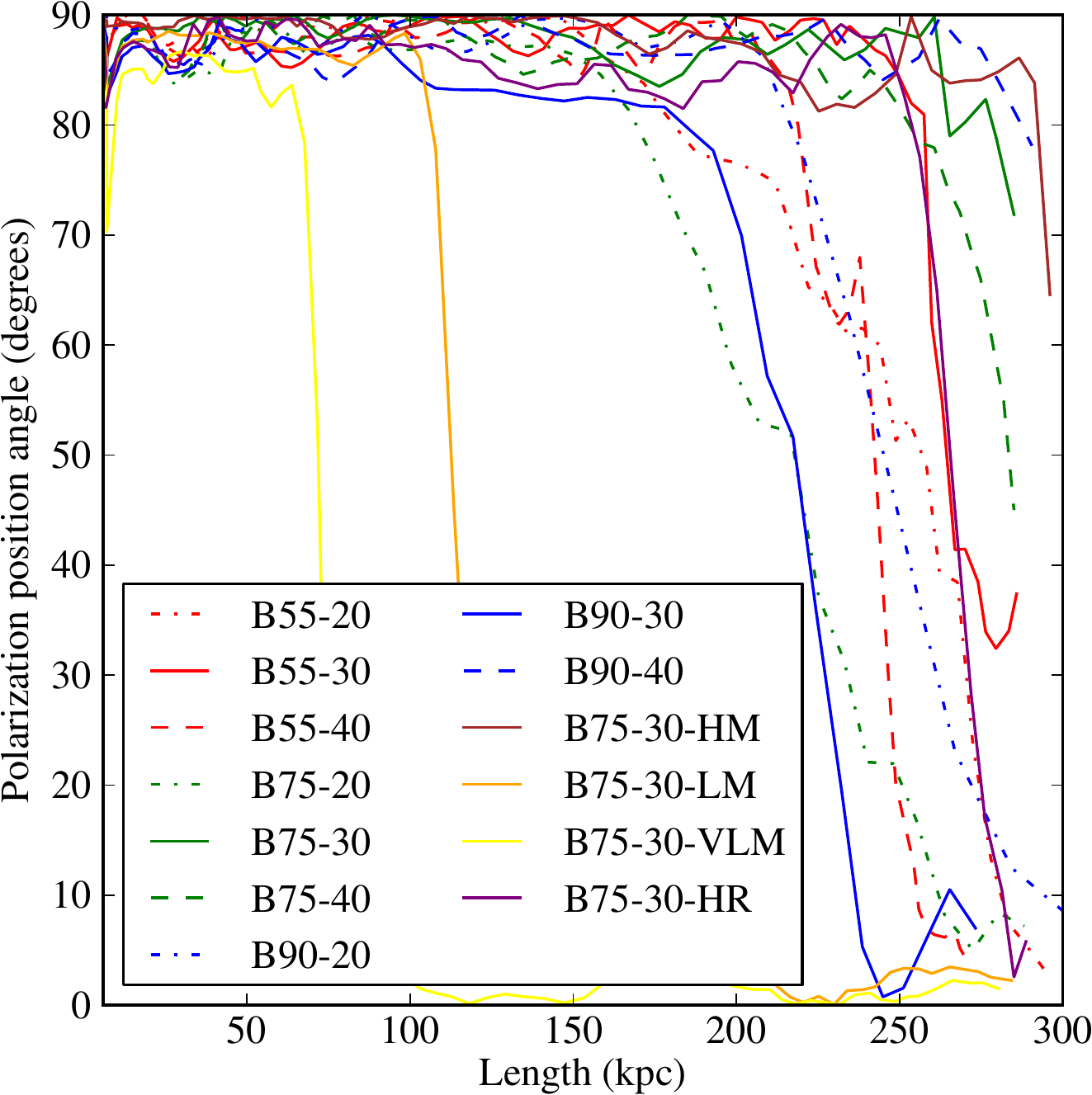}
\includegraphics[width=0.8\linewidth]{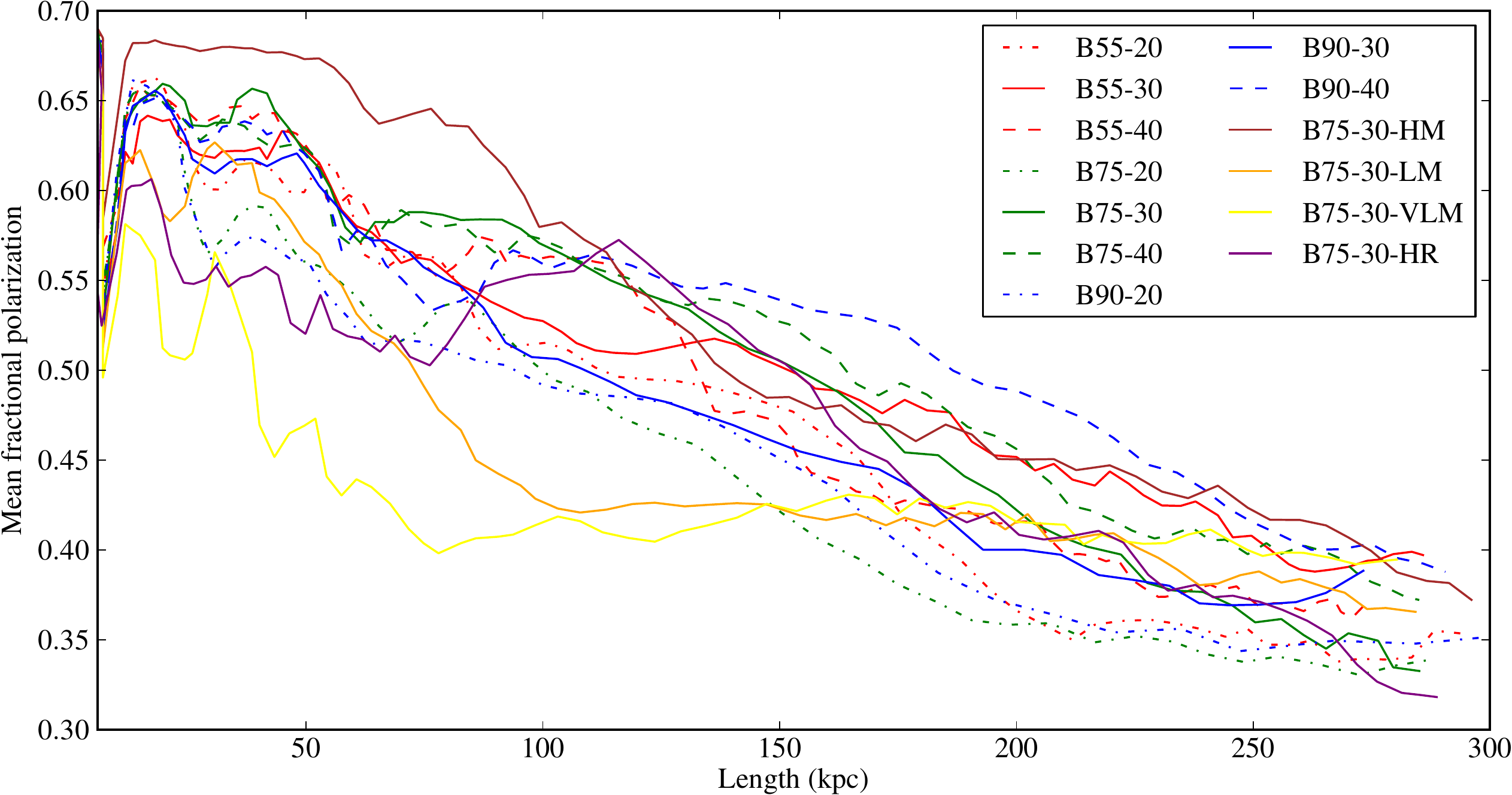}
\caption{Synchrotron emissivity as a function of source length. Top:
  total intensity (luminosity). Middle: integrated fractional
  polarization and position angle (where zero means aligned with the
  jet axis). The horizontal black line shows the
  observed integrated fractional 1.4-GHz polarization for
  steep-spectrum sources of $\sim 2.5$ per cent \citep{Tucci+04}. Bottom: mean fractional polarization at full
  simulation resolution.}
\label{fig:synch-integrated}
\end{figure*}

As in Paper I, we can calculate the integrated radio luminosity to see
how this evolves with time or source length. Here we have the
advantage that we actually know the magnetic field strength as a
function of position, rather than simply assuming some constant
deviation from local equipartition as in Paper I. (In fact, the idea
that the energy density in the field traces that in the particles does
not seem to be a particularly good assumption for these sources, as we
will discuss later.) We therefore compute the Stokes $I$ (total
intensity) synchrotron emissivity for the source making use of the
local magnetic field vector, projecting along the $z$ axis for
simplicity (in the absence of a fully tangled field, synchrotron
emissivity is anisotropic, so we must choose direction from which to
view the source). The Stokes $I$, $Q$ and $U$ emissivities, dropping
physical constants, are then given by
\begin{align}
j_I&=p \left( B_x^2 + B_y^2\right)^{\frac{1}{2}(\alpha - 1)} (B_x^2
+ B_y^2)\\
\label{eq:synchemi}
j_Q&=\mu p \left( B_x^2 + B_y^2\right)^{\frac{1}{2}(\alpha - 1)} (B_x^2
- B_y^2)\\
j_U&=\mu p \left( B_x^2 + B_y^2\right)^{\frac{1}{2}(\alpha - 1)} (2B_xB_y)
\label{eq:synchemu}
\end{align}
where $B_x$, $B_y$ represent the components of the magnetic field
perpendicular to the projection axis, and $p$ is the local thermal
pressure (proportional to the number density of electrons of any given
energy if we assume a fixed power-law electron energy distribution).
$\alpha$ here is the power-law synchrotron spectral index, which we
take to be $\alpha = 0.5$, corresponding to an electron energy index
$p=2$; $\mu$ is the maximum fractional polarization for a given
spectral index, $\mu = (\alpha+1)/(\alpha+5/3) = 0.69$ for $\alpha =
0.5$. These emissivities are then integrated over the lobe volume
  to give a total luminosity. Given the many assumptions that go into
  the conversion to physical units, we present luminosities in
  simulation (i.e. essentially arbitrary) units in what follows; the
  reader is referred to Paper I for discussion of the conversion
  factors.

Results are shown in Fig.\ \ref{fig:synch-integrated}. We see that, as
in Paper I, there is a characteristic rising and falling curve in
total intensity which is followed by most sources; only the object in
the very richest environment (B55-40) does not show any sign of a
downturn in the radio luminosity. Sources in
environments with smaller core radius and larger $\beta$ have the peak
of their emission at smaller linear sizes, while richer environments
produce more luminous radio sources at all times. This gives a strong
reinforcement to the picture, presented in Paper I, in which it is the
environment that determines the track that a
source of a given jet power follows in the power/linear size diagram.
It can also be noted that the high-resolution simulation B75-30-HR
shows trends consistent with its low-resolution counterpart B75-30
within the scatter imposed by the detailed weather in the two
simulated sources; there is no evidence here for effects of numerical
resolution. However, there is clear evidence for the importance of the
input magnetic field strength, with the simulation with the weakest
field, B75-30-VLM, having by far the lowest radio luminosity,
precisely as expected. We emphasise again, as in Paper I, that the
effects of radiative losses on the electron population in the lobes
are not modelled in these simulations, so that they probably
understate the extent of the decline in the radio luminosity at large
lobe lengths.

It is also instructive to plot the integrated fractional polarization
$F_{\rm tot}$, i.e. $\sqrt{Q_{\rm tot}^2 + U_{\rm tot}^2}/I_{\rm
  tot}$. This is the fractional polarization that would be measured if
the source were unresolved, e.g. by a low-resolution radio survey, and
if Faraday rotation effects were negligible; $Q_{\rm tot}$ and $U_{\rm
  tot}$ are the total luminosities at Stokes $Q$ and $U$, obtained by
integrating over the emissivities defined above. $F_{\rm tot}$ is
shown in the middle panel of Fig.\ \ref{fig:synch-integrated} along
with the corresponding absolute value of position angle, where
0$^\circ$ implies a magnetic field direction predominantly along the
$x$-axis (the jet direction) and $90^\circ$ predominantly
perpendicular to it. The point to note here is the very high
fractional polarization at early times. This is probably not realistic
(the fractional polarization of real unresolved sources is generally
low, see e.g. \citealt{Tucci+04}) and stems from the very simple
ordered toroidal field structure in the lobes and jet in the early
parts of the simulations (Fig.\ \ref{fig:bcomp}) and the neglect of
Faraday effects (see below, Section \ref{sec:depol}). The late-time
values of $F_{\rm tot} \la 10$ per cent are probably far more
reasonable; the trend is a result of the increasingly complex magnetic
field structure in the simulations. Support for this picture is
provided by the fact that B75-30-HR is systematically below the other
simulations at early times/short lobe lengths. As the polarization
decreases, the position angle of the integrated polarization shifts
from being perpendicular towards a parallel configuration for most
sources, presumably a result of the increasing dominance of
longitudinal over toroidal field (Fig.\ \ref{fig:bcomp}). The
behaviour of the weak-field simulations B75-30-LM and B75-30-VLM on
this plot are anomalous; it appears that the dynamically irrelevant
fields in these simulations are sheared to give a comparatively
uniform longitudinal configuration with similar polarization
throughout the source.

Finally, we can investigate trends in the mean fractional
polarization, i.e. the mean of the fractional polarization that would
be measured if Stokes $I$, $Q$ and $U$ were measured for pixels at the
full numerical resolution, averaged over all pixels with non-zero
Stokes $I$ (and again projecting along the $z$ axis). This quantity is
what would be measured from high-resolution images, as we will see in
the following section. It shows a very similar trend to the integrated
polarization, and is presumably high for small sources for very
similar reasons, but at large source lengths has only fallen to the
range 30-40 per cent. For the matched-resolution runs, sources in
poorer environments have lower mean fractional polarization. There is
a hint from the B75-30-HR simulation that numerical resolution is
important here -- at late times this simulation shows the smallest
mean fractional polarization, around 30 per cent. We will discuss the
properties of this resolved polarized emission in more detail in the
following section. We note here however that the trends we see in both
integrated and mean fractional polarization with lobe length are
similar to what HE11 saw in their light jet simulations, consistent
with a picture in which these are driven by increasing complexity in
the lobes.

\subsection{Resolved radio emission and polarization}
\label{sec:resolvedradio}

\begin{figure*}
\includegraphics[width=1.0\linewidth]{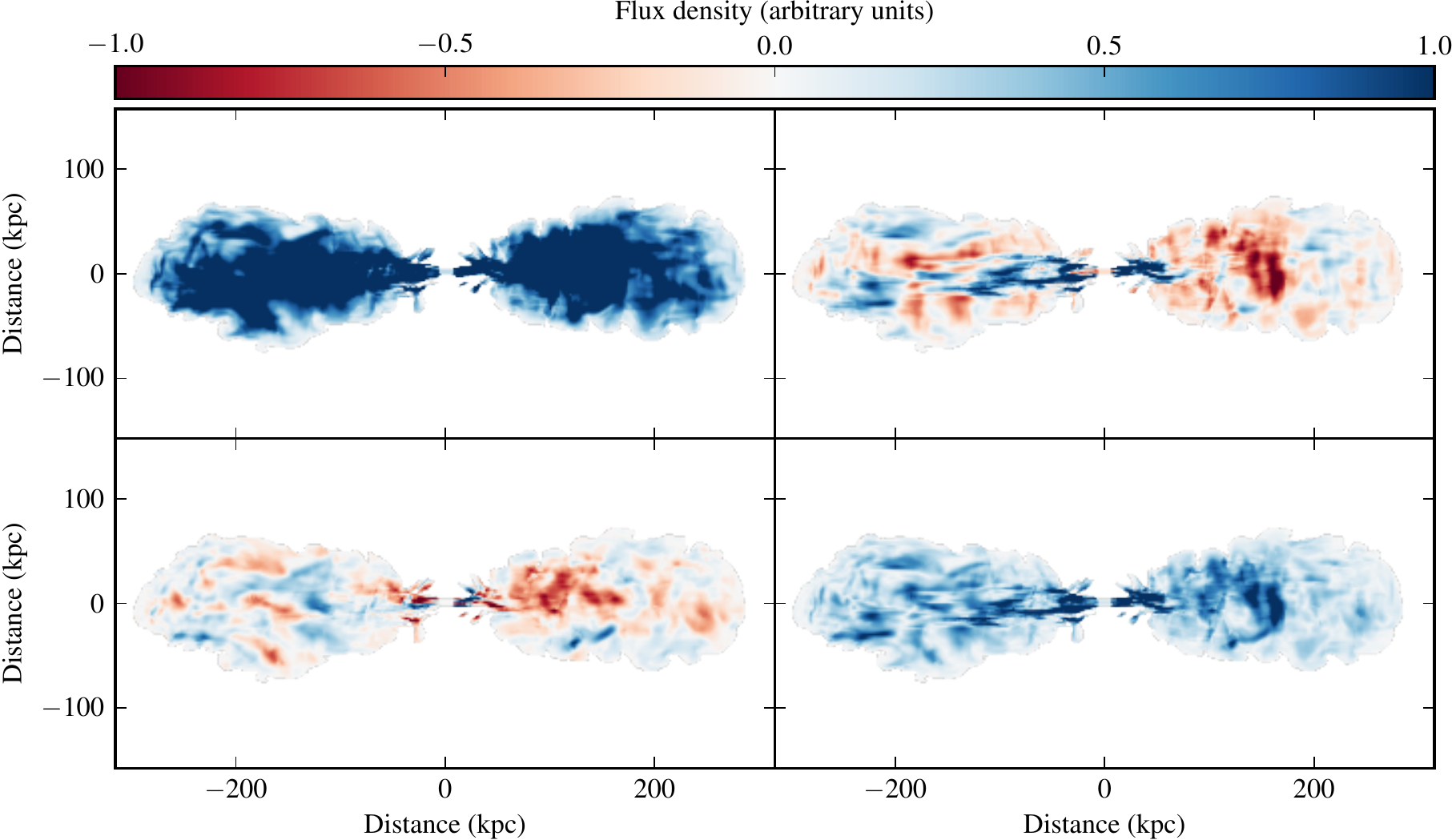}
\vskip 6pt
\includegraphics[width=0.985\linewidth]{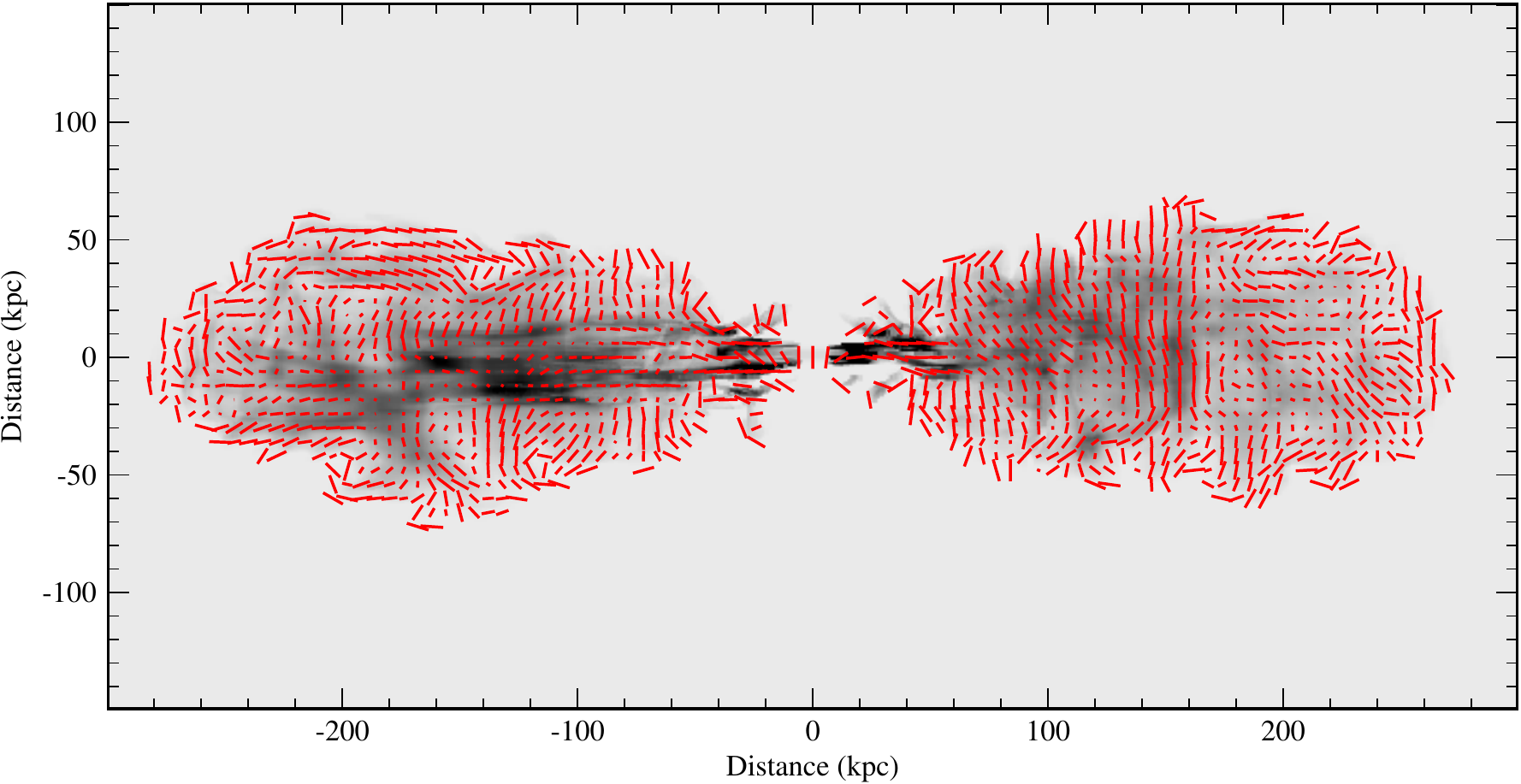}
\caption{Full-resolution synchrotron images of B75-30-HR at $t=200$ Myr. Top row: Stokes $I$ (left) and $Q$ (right). Middle
  row: Stokes $U$ (left) and $P = \sqrt{Q^2+U^2}$ (right), colours as
  for top row. The {\it mean} Stokes $I$ is normalized to unity, so as
  to show faint structure, and this normalization is also applied to
  the other images. Bottom row: the same map (greyscale), with vectors
  whose length indicates fractional polarization and whose direction
  indicates the `magnetic field direction'.}
\label{fig:stokes}
\end{figure*}

One of the main aims of this paper was to extend the results of HE11,
who found relatively high degrees of polarization in MHD simulations
of radio galaxies in realistic environments. Our simulations allow
high-resolution synchrotron imaging of Stokes $I$, $Q$ and $U$. To
visualize synchrotron emission, we take all volume elements in the
simulation with a value of the tracer quantity $>10^{-3}$, and then
compute the emissivity in Stokes $I$, $Q$ and $U$ according to the
relations given above (eqs \ref{eq:synchemi}--\ref{eq:synchemu},
finally projecting along a chosen line of sight to form a
two-dimensional image. Given the discussion of the previous section,
such imaging is most likely to be realistic at late times when complex
magnetic field structures have had a chance to develop.

Fig.\ \ref{fig:stokes} shows an example image, made from B75-30-HR;
note in particular the filamentary structure in Stokes $I$ and the
polarized intensity $P$ (see below, Section \ref{sec:ic} for more on
this) and the patchy, irregular appearance of Stokes $Q$ and $U$, a
result of the complex magnetic field structure in the lobes at late
times. Although the image shows the highest-resolution simulation,
there is no very obvious difference between this and the other,
lower-resolution runs in terms of the general properties of the
images. The contour maps derived from these images
(Fig.\ \ref{fig:stokes}) are characteristic of the results of our
simulations; the `magnetic field direction' ($\theta_m =
\frac{1}{2}\arctan(U/Q) + \pi/2$) is often parallel to the jet
direction at the ends of the source, where backflow is most important,
but perpendicular to the jet axis close to the core, where the
toroidal field structure presumably dominates. It is this change in
the characteristic field direction that presumably gives the low
integrated fractional polarization (see Section
\ref{sec:integratedradio}). Similar behaviour is often seen in the
lobes of real radio galaxies \citep{Hardcastle+97}. Fractional
polarization is generally higher at the edges of the lobes than at the
middle, again as seen by HE11 and also in real radio galaxies. (In
this section we plot only results for sources lying at 90$^\circ$ to
the line of sight; results for smaller angles are qualitatively
consistent with the results for sources in the plane of the sky, as
also noted by HE11, and so are not shown here.)

\begin{figure*}
\includegraphics[width=1.0\linewidth]{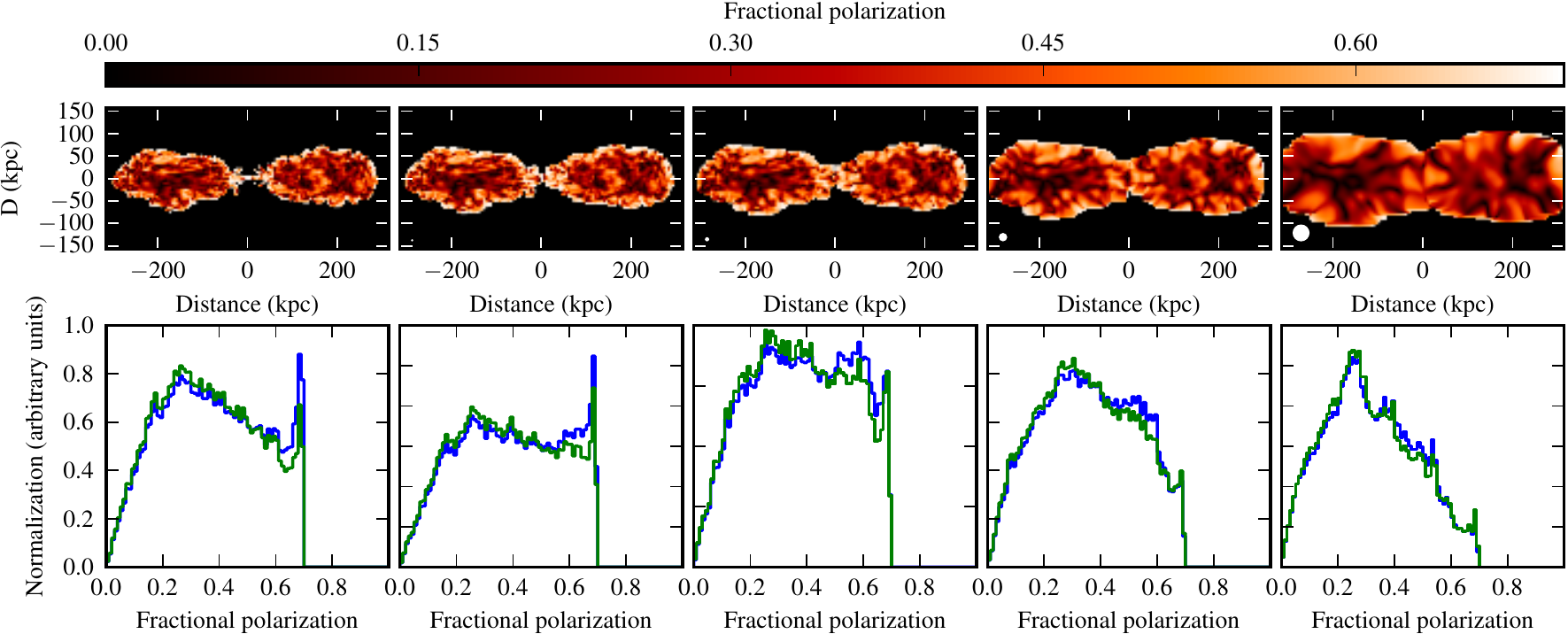}
\caption{The effects of convolution on fractional polarization. Top
  panels show images of fractional polarization from B75-30-HR at
  $t=200$ Myr, while bottom panels show histograms of the
  distribution of fractional polarization with (green) and without
  (blue) the exclusion of the central injection region. From left to
  right: full resolution; convolved with a Gaussian of $\sigma = 1$,
  2, 4, 8 pixels.}
\label{fig:stokesh}
\end{figure*}

It is important to consider how the source would appear if observed
with a finite resolution (as is the case for real sources). This is
illustrated in Fig.\ \ref{fig:stokesh}, where we plot the distribution
of fractional polarization for B75-30-HR (which is representative of
the general source population) at full resolution and
smoothed with increasingly large Gaussians. The general effect of
increasing the convolving beam size is to reduce the mean and maximum
fractional polarization (unsurprisingly, since we have seen above that
the unresolved sources are only marginally polarized). However, even
beams that are an appreciable fraction of the source size give mean
fractional polarizations around 30 per cent for these simulations. Our
{\it unsmoothed} fractional polarization histograms are very
comparable to those obtained by HE11 for their comparable simulations
(light, fast jets at late times), confirming that their results are
not simply the results of the lower numerical resolution of their
simulations. The only important difference is the spike of points with
the maximum possible fractional polarization in our data, which does
not occur in the HE11 simulations. Imaging suggests that this is at
least in part associated with the inner jet (as illustrated by the
green histograms in Fig.\ \ref{fig:stokesh}, which show the
polarization distribution when the injection region is excluded), and
so part of the difference between our simulations and those of HE11
may arise from the different initial conditions for jet polarization
in our simulations. Alternatively, since these very highly polarized
regions are removed by convolution with a moderate-sized Gaussian
beam, they must be small and so may be related to the higher numerical
resolution of our simulations relative to those of HE11. These
fractional polarization histograms are insensitive to the choice of
tracer threshold, as long as it is not too close to unity -- at
thresholds $>0.1$ much of the lobe material disappears and the
polarization seen is mostly that of the jet.

\begin{figure}
\includegraphics[width=\linewidth]{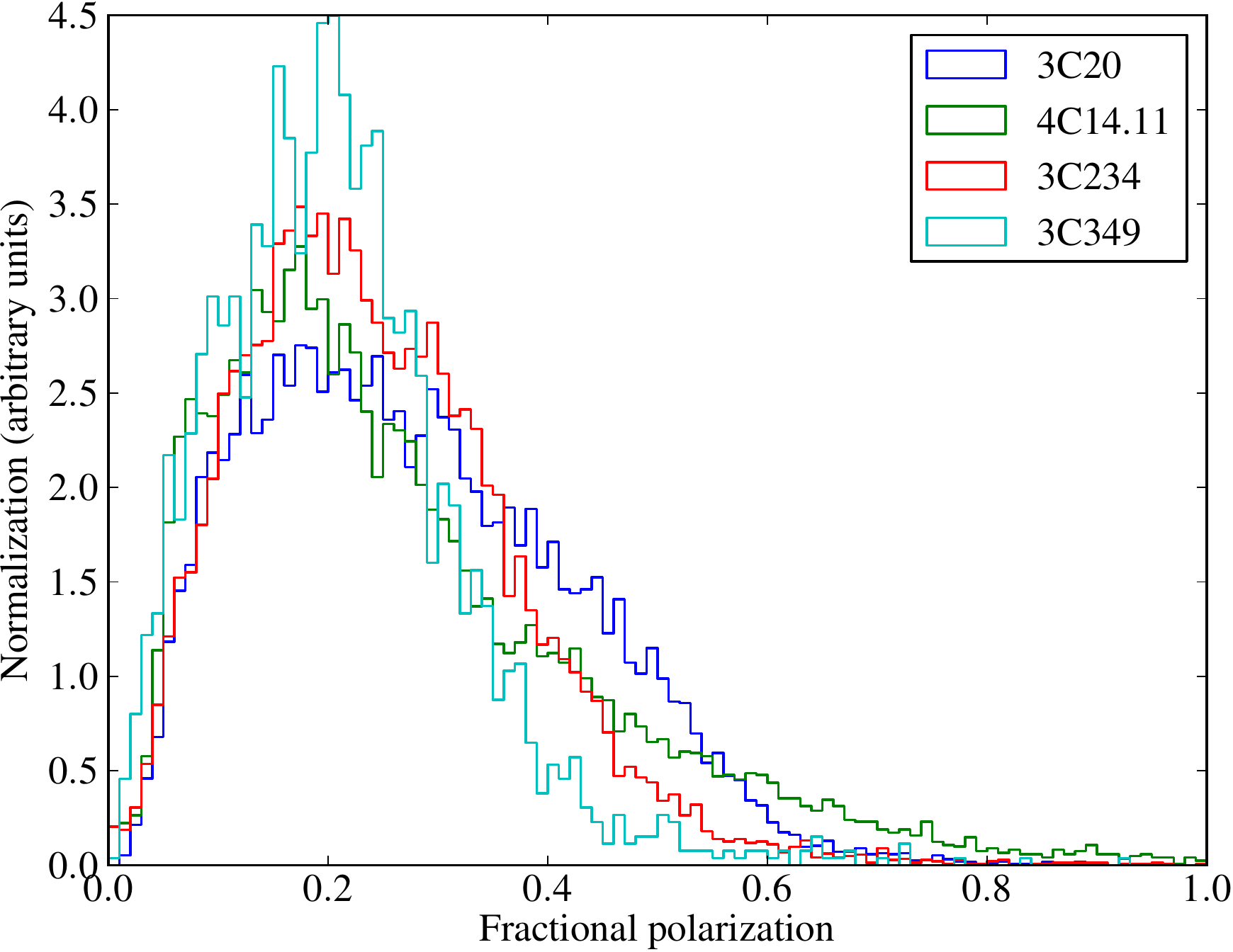}
\caption{Fractional polarization histograms for four radio galaxies
  taken from \cite{Hardcastle+97}.}
\label{fig:realhist}
\end{figure}

How realistic are these distributions of fractional polarization?
Fig.\ \ref{fig:realhist} shows fractional polarization histograms made
at 8.4 GHz (a high enough frequency that rotation measure and
depolarization should not be important) for four large, well-resolved
radio galaxies from \cite{Hardcastle+97}. It is important to note that
these real histograms are affected by noise -- which presumably
accounts for the tail of fractional polarizations $\ga 0.7$ -- as well
as resolution effects. The resolution in the images used here is
between 30 and 50 times the maximum source size, which means that the
histograms should be compared with the final two panels of
Fig.\ \ref{fig:stokesh}. On this basis, the agreement is quite good;
both real and simulated histograms show peaks at fractional
polarization around 20 to 30 per cent with a tail to larger values. A
prediction from our work and that of HE11 is that high-resolution
polarization maps of bright sources will tend to show higher
fractional polarization, perhaps peaking around 40 per cent, and
qualitatively this is indeed seen in higher-resolution maps; however,
for a fair comparison, such maps will need to detect emission from the
whole lobe, which is normally resolved out at the full resolution in
low-sensitiviy maps such as those of \cite{Hardcastle+97}. High-resolution ($LAS/\theta \sim
500$) fractional polarization imaging of Cygnus A
\citep{Perley+Carilli96} does indeed show regions of very high
fractional polarization in both lobes, but this source's polarization
properties are thought to be significantly affected by the rich
intracluster medium, so a direct comparison with our simulations is
difficult, though qualitatively the example of Cygnus A is
encouraging; deep, high-resolution JVLA polarimetry of more typical
radio galaxies is needed to give a better comparison with
observations.

\subsection{Rotation measure and depolarization}
\label{sec:depol}
\begin{figure}
\includegraphics[width=\linewidth]{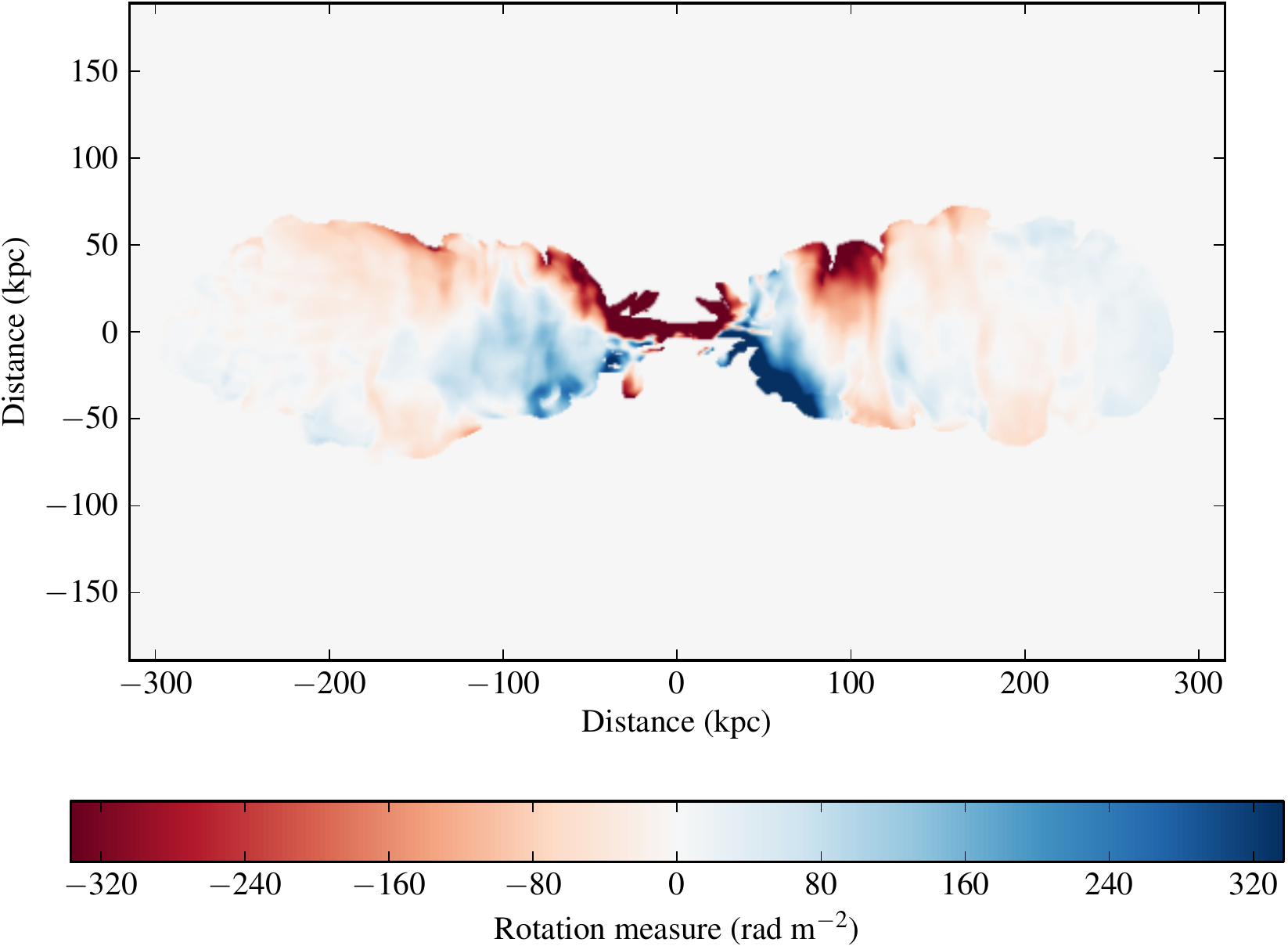}
\caption{Example rotation measure visualization: B75-30-HR in the
  plane of the sky at $t=200$ Myr.}
\label{fig:rm}
\end{figure}

\begin{figure*}
\includegraphics[width=0.8\linewidth]{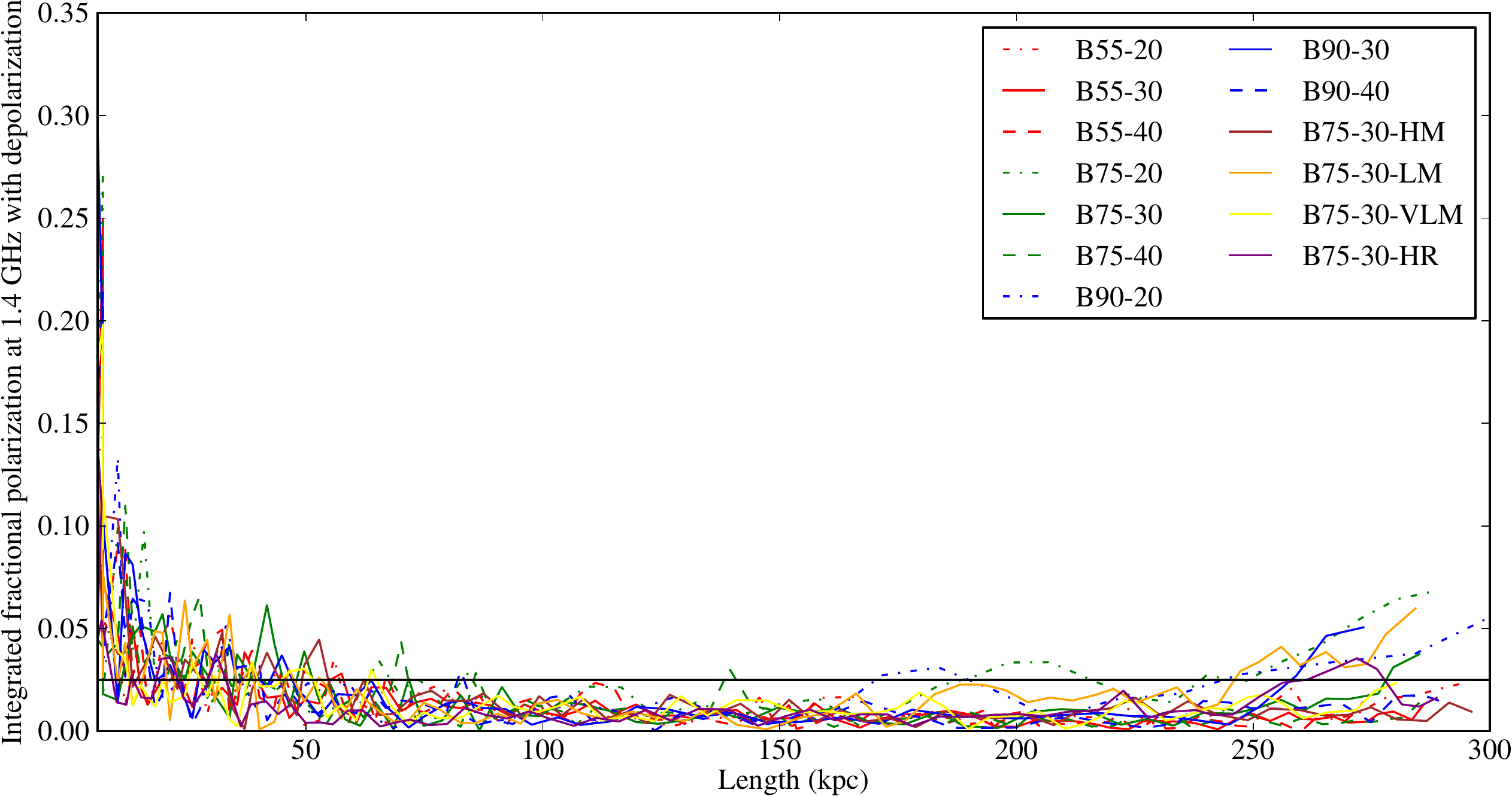}
\includegraphics[width=0.8\linewidth]{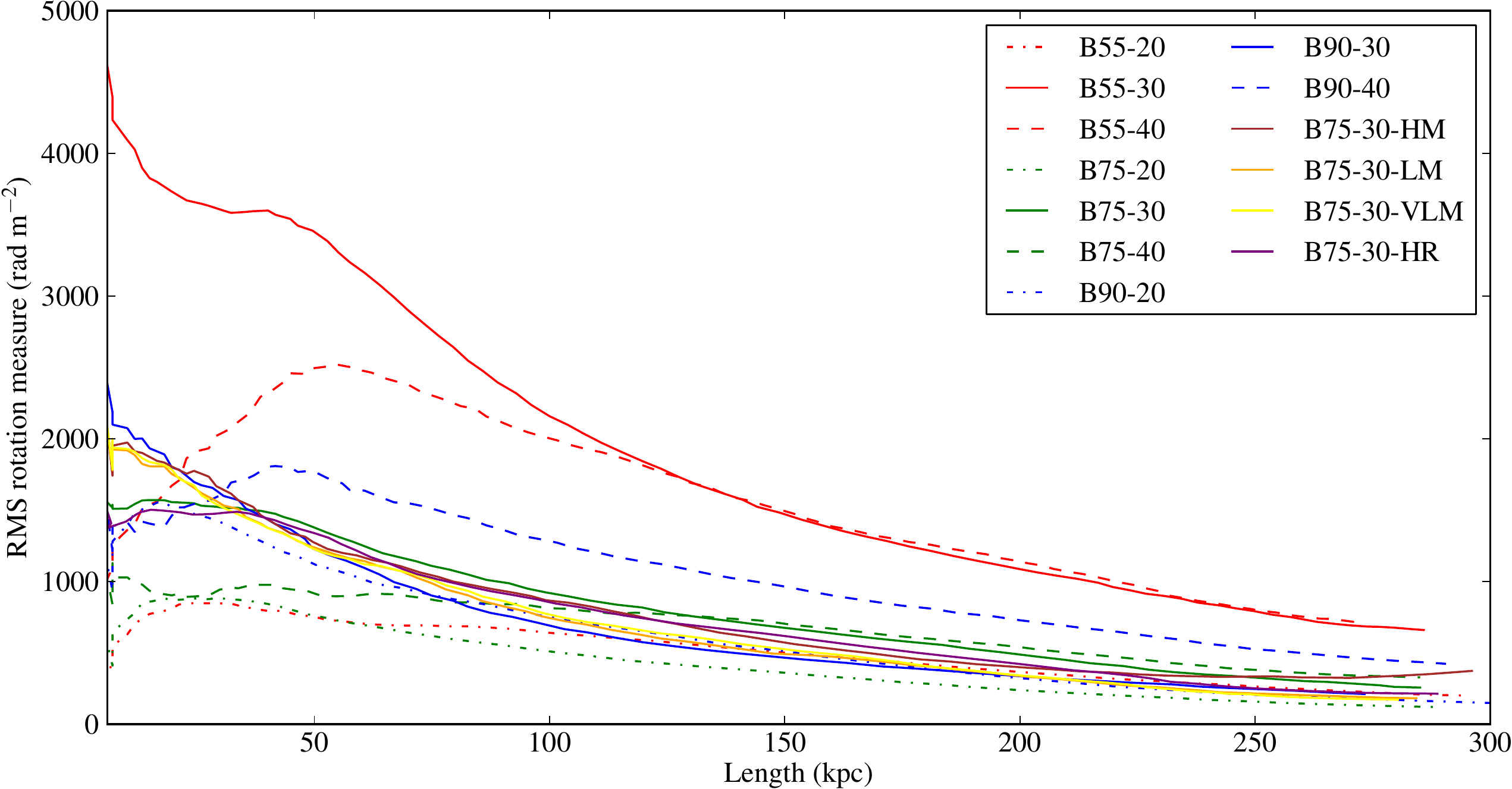}
\includegraphics[width=0.8\linewidth]{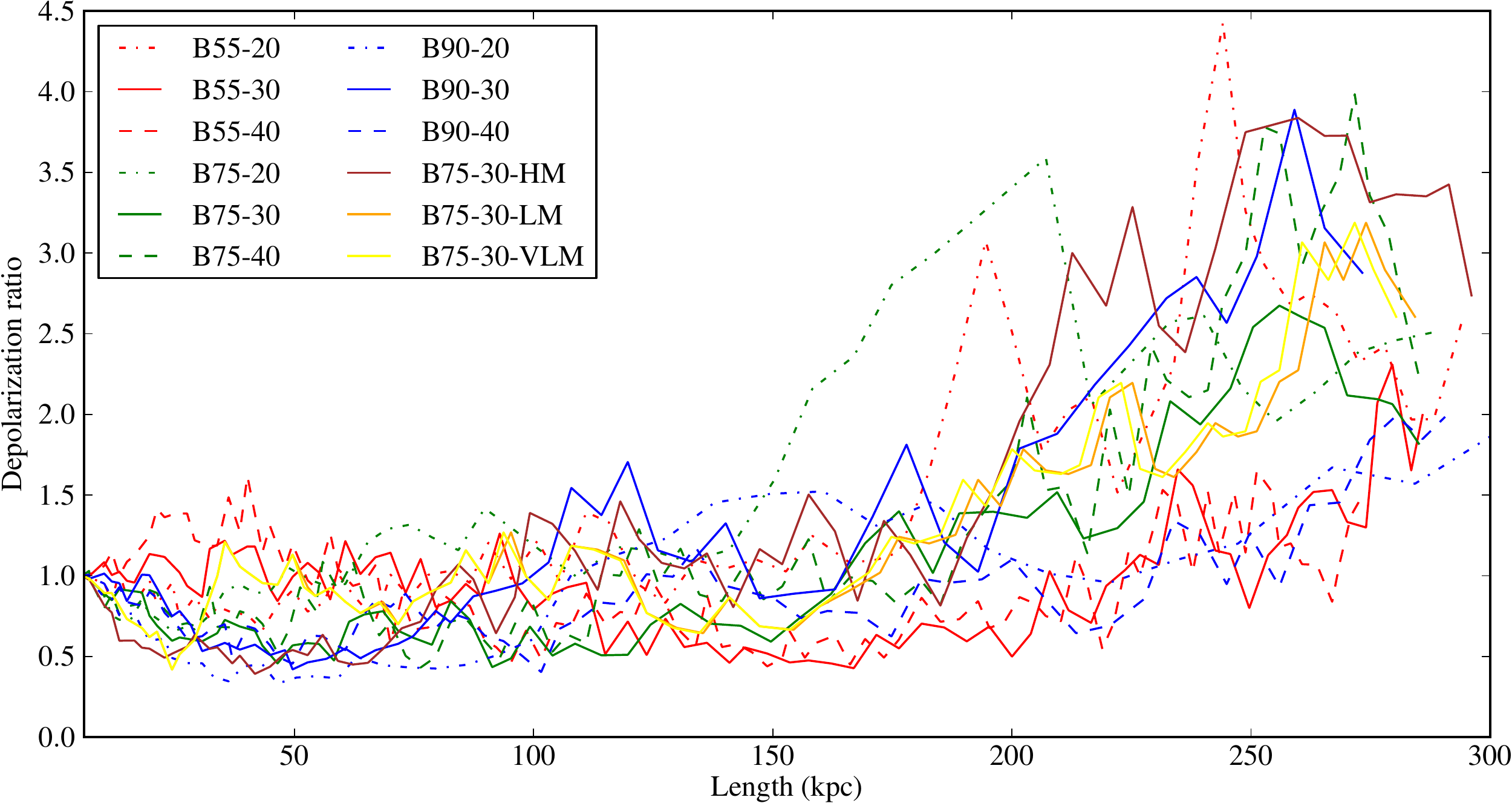}
\caption{Effects of Faraday rotation as a function of lobe length. Top panel: integrated fractional
polarization at 1.4 GHz. The solid black line shows the observed
integrated 1.4-GHz fractional polarization of unresolved
steep-spectrum sources \citep{Tucci+04}. Middle panel: RMS value of the rotation
measure in RM images. Bottom panel: depolarization ratio for lobes at
45$^\circ$ to the line of sight (the Laing-Garrington effect).
B75-30-HR is omitted from the bottom panel as a uniform path length
through the shocked external medium cannot be achieved at late times.}
\label{fig:depol}
\end{figure*}

A magnetised intracluster medium gives rise to Faraday rotation which
(if wholly or partially unresolved) will lead to depolarization
\citep{Burn66,Laing84}. Here we are only concerned with external
depolarization, the effect of the medium between the radio lobe and
the observer, as there is little direct evidence for internal
depolarization in FRII lobes, and we cannot in any case adequately
model the microphysics of thermal entrainment in FRIIs.

The rotation measure ($RM$) is defined as
\begin{equation}
RM = C\int n_{\rm th} B_{\parallel}{\rm d} z\ {\rm rad}\ {\rm m}^{-2}
\label{eq:rm}
\end{equation}
where $C$ is a constant which in physical units is $2.62 \times
10^{-13}$ T$^{-1}$, $n_{\rm th}$ is the thermal density and $B_{\parallel}$ is
the component of magnetic field parallel to the line of sight. The
actual angle of Faraday rotation in radians is then given by $\phi =
RM\lambda^2$, where $\lambda$ is the wavelength of observation. As we
have chosen the values of our field strength and density to match
observed properties of clusters of this type, we can easily compute
$RM$, and so $\phi$, in physical units.

We visualize rotation measure by integrating as specified by
eq.\ \ref{eq:rm} from an outer radius defined by the edge of the
computational volume along the $x$ axis (320 kpc) along the line of
sight until the near edge of the lobes is reached (defined as usual
using tracer values $>10^{-3}$). This allows maps of RM to be
constructed (Fig.\ \ref{fig:rm}). \cite{Huarte-Espinosa+11} have
discussed such synthetic RM images in detail and it is not our purpose
in this paper to repeat their analysis. However, we are exploring a
rather different regime of cluster parameter space -- our sources are
physically larger at the grid edges, considerably lower in jet power,
and inhabit clusters/groups with smaller core radii. As a result, we
would in general expect lower values of the rotation measure at late
times, which is indeed what we observe. A characteristic size
  scale of the structures in RM of order tens of kpc is observed,
  which, given that the largest scale of the power spectrum of the
  magnetic field is 40--80 kpc, is consistent with the estimates of
  the coherence length for RM in the undisturbed ICM given by
  \cite{Cho+Ryu09}. We also note in some, though not all, simulated
sources a tendency for a systematic side-to-side asymmetry in RM,
which we attribute to the predominantly toroidal field of the radio
galaxy `leaking' (either via small-scale mixing at the boundary
  layer or numerically) into the shocked material; this is not so
obvious in the maps of \cite{Huarte-Espinosa+11}, though there are
hints that it may be present in the lightest-jet simulations which are
the best match to those we use here. This effect is independent of
tracer threshold and so is not simply an effect of lobe material which
is not being classified as such; it appears to be important where
  the edge of the lobe becomes K-H unstable, giving rise to fine
  structure and numerical mixing at the lobe boundary. However, as the
  effect can be removed by cutting off the RM integration within a few
  volume elements of the lobe boundary without significantly affecting
  the range or dispersion of the RM, we ignore it in the analysis that
  follows.

Rotation measure causes depolarization if it is unresolved or
partially resolved. It has a particularly strong effect on integrated
polarization, as defined in Section \ref{sec:integratedradio}.
Fig.\ \ref{fig:depol} (top panel) shows the results of `observing' at
1.4 GHz (chosen as it is the frequency of the FIRST and NVSS surveys)
taking the effects of rotation measure into account. The fractional
polarization is reduced to low levels, almost independent of source
size, by the depolarization from the intracluster medium; we do not
expect in practice to see the high levels of integrated polarization
shown in Fig.\ \ref{fig:synch-integrated}. The integrated fractional
polarization plotted is now consistent with the observational value.
Fig.\ \ref{fig:depol} also shows the RMS value of rotation measure
seen by the lobes (at 90 degrees to the line of sight) and, as
expected, we see that this generally decreases with time and is larger
for sources in denser environments. The non-negligible values of
$<RM^2>^{1/2}$ even for large sources means that we can expect
substantial rotation/depolarization in low-frequency polarization
observations of such objects.

Finally, we can estimate the magnitude of the Laing-Garrington effect
\citep{Laing88,Garrington+88,Garrington+91} for our sources. The
Laing-Garrington effect is the tendency of the jetted (therefore
nearer) lobe to be less depolarized than the counterjet (further)
lobe, as a result of the longer path length to the further lobe
through the depolarizing medium. Although X-ray observations are
consistent with the observed polarization properties of radio sources
in a few cases \citep{Belsole+07} there has been no systematic
observational comparison between polarization statistics and
group/cluster properties; our simulations allow us to make some
predictions. We simulate the observations described by
\cite{Garrington+91} by considering lobes at 45$^\circ$ to the line of
sight, generating and applying the RM map at two frequencies of 1.4
and 5 GHz, and then convolving with a Gaussian designed to give
roughly 15 beam widths across the source (matching the observations of
\citealt{Garrington+91}). The resolved fractional
polarization is then measured for each lobe at each frequency, the
depolarization ($DP = f_{\rm 1.4}/f_{5}$ is calculated for the two
lobes, and finally we compute the depolarization ratio, $DPR =
DP_j/DP_{cj}$, where the `jet' and `counterjet' lobes are simply taken
to be those closer to and further away from the observer. $DPR$ by
this definition should be larger than 1 to be in the sense expected
from the Laing-Garrington effect. The results are plotted in
Fig.\ \ref{fig:depol} and it can be seen that a Laing-Garrington
effect at these frequencies is indeed expected even for these
relatively poor environments, especially for large lobe lengths. The
magnitude of $DPR$ is very comparable to what is found in
observations. Two other points are worth noting about this plot; the
comparatively large scatter and the lack of a strong Laing-Garrington
signal at small lobe lengths. The scatter is perhaps not surprising,
as with each time step the lobe is advancing into a new part of the
external medium with potentially different magnetic field structure
(recall the turbulent initial magnetic field conditions) but shows
that the absolute magnitude of the Laing-Garrington effect in
individual sources will be hard to use to infer physical conditions
for the intracluster medium; resolved depolarization/RM imaging will
in general be far more robust for these purposes. The lack of a strong
effect at small lobe lengths is perhaps surprising given that
\cite{Garrington+91} report stronger Laing-Garrington signals for
sources of small angular size, but of course small angular sizes also
tend to be associated with more distant, more luminous objects
plausibly in richer X-ray environments (cf.\ \cite{Ineson+13}) which
might then have higher DPR for a given physical size.
Additionally, there may be some difficulty in imaging the
Laing-Garrington effect in our simulations at small angles to the line
of sight due to the large source axial ratios at early times.

\subsection{Inverse-Compton visualization}
\label{sec:ic}

\begin{figure*}
\includegraphics[width=1.0\linewidth]{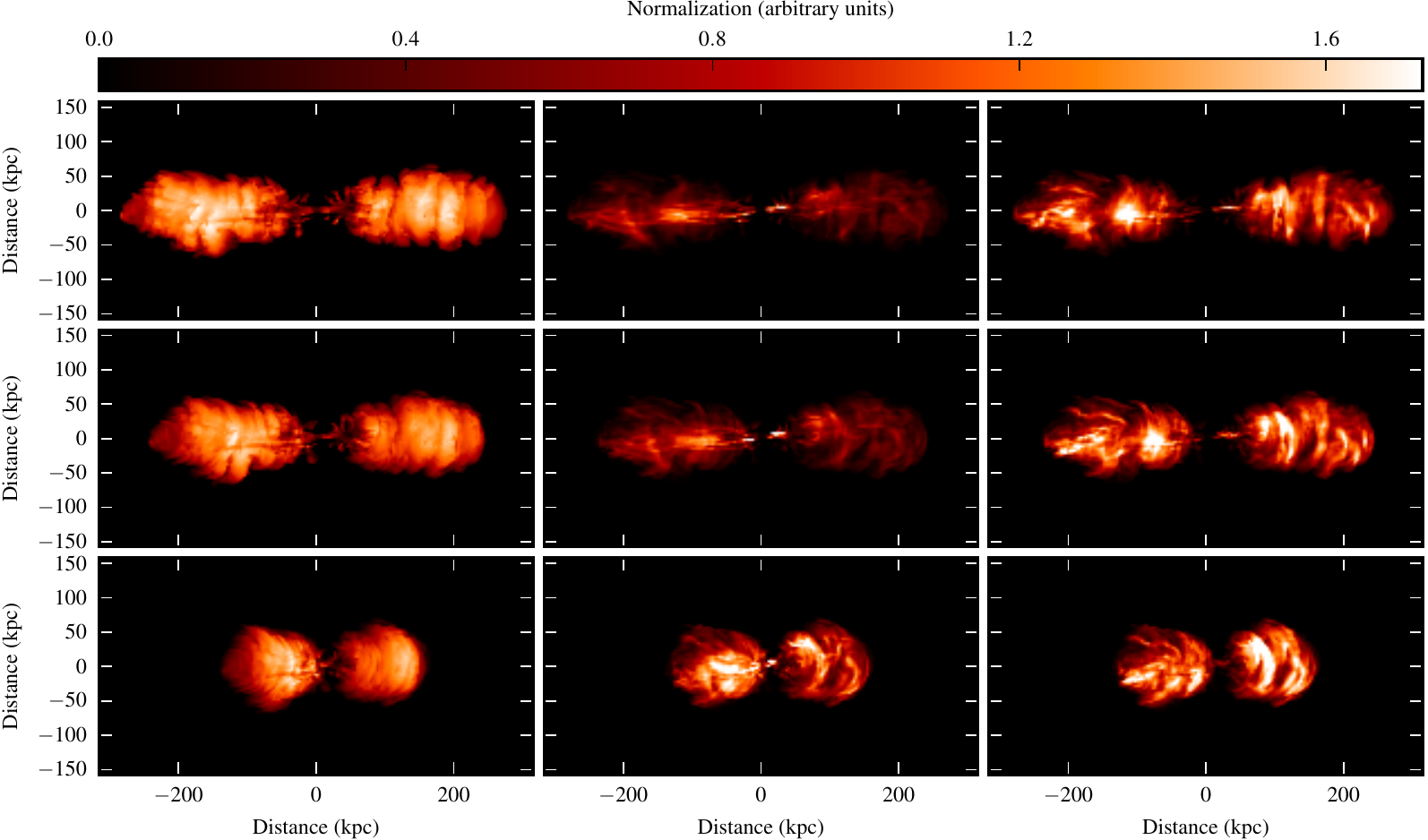}
\caption{Inverse-Compton and synchrotron visualizations of B75-30-HR
  at $t=200$ Myr.
  From top to bottom, the simulation is at 0$^\circ$, 30$^\circ$ and
  60$^\circ$ to the plane of the sky. Left panel: inverse-Compton;
  centre panel: synchrotron; right panel: synchrotron divided by
  inverse-Compton. The colour scale is linear with arbitrary units.}
\label{fig:ic}
\end{figure*}

We can very easily visualize inverse-Compton scattering of the CMB in
these models by simply integrating particle (thermal) pressure along
the line of sight (as before we require values of the tracer
$>10^{-3}$ for this analysis). The CMB is an isotropic photon field
and is scattered by low-energy electrons whose number density will be
strictly correlated with the local value of pressure (unless there is
an energetically dominant, spatially varying proton population: see
discussion in Section \ref{sec:critik}). By doing this we can compare
with the few existing results on the resolved relationship between
inverse-Compton and synchrotron emission
\citep{Hardcastle+Croston05,Goodger+08}.

Results of this analysis for B75-30-HR are shown in
Fig.\ \ref{fig:ic}. Again, while we use the high-resolution simulation
for display purposes, these images are representative of all of our
simulations. Here we plot three angles to the line of sight to show
that our conclusions are not strongly affected by viewing angle. We
see that the simulated inverse-Compton emission is much more uniform
across the lobe than the synchrotron emission, exactly as seen in real
sources such as Pictor A \citep{Hardcastle+Croston05}: consequently,
the ratio of synchrotron to inverse-Compton, which essentially
represents the pressure-weighted projection of the magnetic field
terms in eq.\ \ref{eq:synchemi}, exhibits strong structure. Clearly
this arises because the magnetic field (or more precisely the energy
density in the components perpendicular to the line of sight, see
eq.\ \ref{eq:synchemi}) is very much more intermittent than the
pressure. \cite{Tregillis+04}, who also simulated light, fast jets,
obtained similar results, in the sense that the scatter in their
synchrotron flux was very much larger than the scatter in their
inverse-Compton flux, and \cite{Gaibler08} finds a similarly smooth
appearance in inverse-Compton visualizations of axisymmetric lobes, so
we suggest that this is a general feature of MHD models where the
field is not dynamically important. We would argue therefore that
these simulations strongly support the arguments of e.g.
\cite{Hardcastle+Croston05} and \cite{Hardcastle13}: filamentary
structures seen in total intensity synchrotron radiation are primarily
tracing intermittency of the magnetic field and can be used to derive
information on the magnetic field power spectrum.

It can readily be seen that the {\it total} inverse-Compton emissivity
in these models (and for this visualization) scales with $\int P{\rm
  d}V$ and so essentially traces the total thermal energy in the
lobes, which increases linearly with time (Section
\ref{sec:energetics}). Thus we expect large, old sources in poor
environments where the synchrotron emission has passed its peak
(Section \ref{sec:integratedradio}) to
have larger ratios of inverse-Compton to low-frequency synchrotron
emission, irrespective of the ratio of magnetic to particle energy of
the lobe (which is in fact roughly constant in these simulations at
all but the earliest times: Fig.\ \ref{fig:bfield}).

\section{Discussions and conclusions}
\label{sec:discussion}

\subsection{Critique}
\label{sec:critik}

We should begin by acknowledging the ways in which our simulations are
{\it not} representative of real radio galaxies, since this helps us
to understand the extent to which our results tell us about real
observations.

Although we have tuned the simulation parameters to match those of
real sources as closely as possible within our numerical limitations,
we are almost certainly some way off reality in the lobes. In FRIIs
there is no evidence for an energetically dominant (thermal or
non-thermal) proton population \citep{Croston+05-2}. The absence of
thermal content in the lobes means that the effective matter density
in lobes is very low indeed, $\sim 3p/c^2$. For $p \sim 10^{-11}$ Pa
(the pressure in the centre of our environments) the matter density
would be equivalent to 0.2 protons m$^{-3}$, or about a factor $10^5$
below the ambient central density; our lobes have a density contrast
at most a little above $10^4$ (Fig.\ \ref{fig:slices}). The equation of state is
also wrong in the lobes -- even the observed density contrast implies
temperatures $10^4$ times higher than ambient for rough pressure
balance, and thus $kT = 20$ MeV, implying relativistic electrons and a
ratio of specific heats of $\frac{4}{3}$. And, possibly most
importantly, the speed of the jet is high enough ($0.24c$) that
relativistic effects would ideally be taken account of in the
modelling, but at the same time too low to match observations, which
require a minimum of $\sim 0.5c$ \citep{Mullin+Hardcastle09} and maybe $>0.99c$
\citep[][and references therein]{Konar+Hardcastle13}. We could only accommodate such high speeds
while maintaining the jet energetics and high density contrast by
reducing the input jet radius, which would improve the realism of the
simulations (a jet radius of a few kpc is certainly too large on the
smallest scales we probe) but this would require higher numerical
resolution for adequate modelling, as well as the use of relativistic
codes.

A consequence of our choice to limit the speed is that the jet
is only mildly supersonic with respect to its internal sound speed.
Consequently, there is no strong jet termination shock, and our
simulations do not form bright hotspots, though there are weak
transient shock structures within the lobes at various times. In fact,
synchrotron visualizations of the lobes (Fig.\ \ref{fig:stokes}) tend to
show structures related to the jet, i.e. to have an FRI-like or
intermediate FRI/II structure, not unlike some real radio sources
(e.g. Hercules A, \citealt{Gizani+Leahy03}) but certainly not typical
of the FRII population. Since the lobe dynamics are dependent on
  ram pressure balance at the front of the lobe and on the lobe
  temperature and density, none of which depend explicitly on the jet
  internal Mach number, we would argue that this does not affect our
ability to understand lobe dynamics and energetics, but it may affect
our conclusions about observability.

Turning to physics that is absent altogether, the most important
missing piece of the picture is particle acceleration and loss, which
is well known to be important in real radio galaxies -- we have the
capability to map this in great detail in observations with the
current generation of radio telescopes \citep{Harwood+13}. Including
acceleration processes would certainly increase the visibility of
hotspots (see the simulations of e.g.\ \cite{Tregillis+01} where these
processes are taken into account, or those of \cite{Krause+12} where a
shock tracer is used to infer high-frequency emissivity) while
including radiative losses would also affect our modelling of the
evolution of lobe radio luminosity (Section \ref{sec:integratedradio})
although, again, neither of them would affect the dynamics
significantly (the total energy radiated away by a radio galaxy in its
lifetime is only a small fraction of what is stored in the lobes). Of
course, in order to model acceleration processes usefully, we would
need more strongly supersonic speeds in the jet, as discussed above,
together with some prescription for particle acceleration at internal
shocks. Cooling in the intracluster medium is also not taken into
account, though this is less important for the simulated media we have
used and the lifetimes of the sources ($\sim 10^8$ years).

\subsection{What we have learnt}

We can now summarize what we think these new simulations tell us,
bearing in mind the limitations imposed by the discussion of the
previous section.

\begin{itemize}
\item The broad dynamical conclusions of Paper I, as expected, are
  reproduced (Section \ref{sec:dynamics}): lobes of this type are not
  self-similar and come into rough pressure equilibrium with their
  surroundings. Lobes are driven away from the centre of the cluster
  environment at late times by the return of the dense central cluster
  material. As the lobes are effectively vacuum, the fact that we have
  not achieved realistic values of the density contrast probably will
  not significantly affect this conclusion.
\item One of the key conclusions of Paper I, that the models imply
  rough energy equipartition between the lobes and the shocked
  material surrounding them, is robust (Section \ref{sec:energetics})
  -- it is hard to see how this could depend on the details of the jet
  dynamics or the density contrast, although the lobe equation of
  state would be expected to affect it to some extent and, as
    noted above, we are not yet simulating jets of the correct speed
    or internal Mach number. It almost
  certainly does depend on the lobes coming into pressure balance, and
  would not be expected to be true in the early, strong-shock regime
  (cf.\ \cite{Cielo+14}).
\item Another conclusion of Paper I, that the evolution of radio
  luminosity should be strongly affected by environment, is reproduced
  here (Section \ref{sec:integratedradio}). Our tracks in the
  power/linear size diagram (Fig.\ \ref{fig:synch-integrated}) are
  probably more realistic than those of Paper I because of our more
  realistic source dynamics at early times and because of the
  inclusion of magnetic fields: they still show close to an order of
  magnitude difference, for a given length, between sources of
  identical intrinsic jet power, which, as discussed in Paper I,
  renders radio-luminosity-based estimates of jet kinetic power $Q$
  very uncertain. Because the integrated luminosity depends largely on
  the lobe volume and pressure, this conclusion cannot be sensitive to
  details of the jet dynamics in the regime where the lobe is in rough
  pressure balance. Of course, loss and acceleration processes would
  significantly change the shape of these curves at high frequency,
  but they would be very unlikely to do so in such a way as to bring
  them into systematically closer agreement.
\item Lobes with a magnetic field that is not dynamically dominant,
  given any reasonable initial field configuration,
  rather naturally evolve towards a state where the energies in the
  toroidal and longitudinal field components are of the same order of
  magnitude, arising because of the balance between the growth of the
  toroidal field through the `toroidal stretching process'
  (\citealt{Matthews+Scheuer90,Gaibler+09}; HE11) and the effects of
  shearing of the toroidal field from jet and backflow (Section
  \ref{sec:general}, \ref{sec:energetics}) which together give rise to a
  predominantly longitudinal component with many field reversals. (The
  growth of this disordered longitudinal component is presumably
  limited by reconnection on the smallest scales, though our results
  do not appear to be strongly-resolution-dependent.) The presence of these two
  significant field components gives rise to complex polarization
  structure at all late times in the simulations (i.e. at all times
  when large-scale bulk motions within the lobes are capable of
  operating), as also seen by HE11, so long as the initial field is
  not too weak. These processes can only be
  studied in 3D. Here the limitations of our simulations are related
  to the internal bulk speeds and densities and our assumptions about
  the initial field configuration and strength, but, so long as the
  field does not become dynamically dominant, it is hard to see how
  these could make a qualitative difference.
\item The complex field configuration that we observe (and that has
  been observed previously by e.g. \cite{Tregillis+04} and
  HE11 gives rise to a number of important
  features of the total and polarized emission. First of all, we note
  that filamentary structure seen in total intensity (e.g.
  Fig.\ \ref{fig:stokes}) is almost entirely due to intermittency in
  the magnetic field. In fact, we can take out the variation in the
  line-of-sight electron pressure altogether (Fig.\ \ref{fig:ic}) and
  still see all the filamentation, as HE11 also remark. This is in
  fact very natural in a situation where the magnetic field is not
  dynamically dominant -- pressure variations in the electrons will be
  washed out on a sound-crossing time (for the high internal sound
  speed of the lobes) while magnetic filaments can persist and grow.
  An observational consequence of this is that inverse-Compton
  visualizations show a much smoother structure than synchrotron
  (Section \ref{sec:ic}).
  There is some direct evidence from observation that this is the case
  in real radio galaxies \citep{Hardcastle+Croston05}.
\item Turning to polarized emission, we see that the complex
  polarization structure at late times implies low integrated
  polarization (Section \ref{sec:integratedradio}) as observed.
  (Conversely, the high integrated fractional polarization implied by
  our simulations at early times, before the lobe internal dynamics
  have had any effect on the field, is probably not realistic,
  although, as discussed in Section \ref{sec:depol}, it would probably
  be washed out by Faraday effects in any case.) However, the fact
  that large-scale magnetic field structures have a strong effect on
  emissivity means that there are expected to be regions of the lobe
  with high fractional polarization (Section \ref{sec:resolvedradio}).
  We have compared observed and simulated fractional polarization for
  the first time, finding good qualitative agreement, but we predict
  that as high-resolution, sensitive polarization observations of
  lobes become available, the maximum polarization seen will approach
  the theoretical maximum, with a complex relationship between
  structures in total intensity and polarization
  (Fig.\ \ref{fig:stokes}). These results confirm the work of HE11
  with numerical resolution up to a factor 3 higher.
\item Finally, we have briefly investigated resolved depolarization
  effects of the magnetized external medium on the lobes, following
  \cite{Huarte-Espinosa+11}. We reproduce the Laing-Garrington effect
  in a numerical model, the first time that this has been done, and so
  demonstrate that the physical conditions in even a moderate-density
  environment are capable of producing a significant Laing-Garrington
  effect at GHz frequencies (Section \ref{sec:depol}). The ability to
  model depolarization effects of this type will become increasingly
  important with the advent of low-frequency polarimetry. We caution
  that some of the quantitative detail discussed here (e.g. the way in
  which the Laing-Garrington effect only becomes important with source
  sizes comparable to the core radius) depends on the lobe dynamics at
  small scales {\it and} our assumptions about the cluster magnetic
  field strength and power spectrum, which may not be realistic.
\end{itemize}

In future work, we plan to verify these conclusions for more realistic
(i.e. not spherically symmetric) cluster environments and include the
effects of particle acceleration and loss in our modelling, with the
ultimate aim of generating recipes by which jet power and source
environment can be inferred accurately from multi-frequency, resolved,
full-polarization images of radio galaxies.

\section*{Acknowledgements}

We thank Mart\'\i n Huarte-Espinosa and Judith Croston for helpful
discussions on aspects of the paper. MGHK acknowledges support by the
cluster of excellence ‘Origin and Structure of the Universe’
(http://www.universe-cluster.de/). We thank an anonymous referee
  for a careful and constructive reading of the paper.

This work has made use of the University of Hertfordshire Science and
Technology Research Institute high-performance computing facility.
This research made use of APLpy, an open-source plotting package for
Python hosted at http://aplpy.github.com (enhancements to allow
plotting of polarization vectors may be obtained from
http://github.com/mhardcastle/aplpy/tree/show-vectors).

\bibliographystyle{mn2e}
\renewcommand{\refname}{REFERENCES}
\setlength{\bibhang}{2.0em}
\setlength\labelwidth{0.0em}
\bibliography{../bib/mjh,../bib/cards}

\end{document}